\documentclass[useAMS,usenatbib]{mn2e}
\usepackage{graphicx}
\usepackage{natbib}
\usepackage{lscape}
\usepackage{longtable}
\usepackage{subfig}
 \usepackage{times}

\newcommand{\apj}{ApJ}

\newcommand{\CIV}{C\,{\sc iv}}
\newcommand{\MgII}{Mg\,{\sc ii}}

\newcommand{\AlIII}{Al\,{\sc iii}}

\newcommand{\CIII}{C\,{\sc iii}]}
\newcommand{\FeII}{Fe\,{\sc ii}}

\newcommand{\NV}{N\,{\sc v}}

\newcommand{\SiIV}{Si\,{\sc iv}}

\newcommand\ion[2]{#1$\;${\small\rmfamily\@Roman{#2}}\relax}%

\def\lsim{\lower0.3em\hbox{$\,\buildrel <\over\sim\,$}}
\def\gsim{\lower0.3em\hbox{$\,\buildrel >\over\sim\,$}}

\title[Spectropolarimetry of BALQSOs]{The orientation and polarization of broad absorption line quasars\thanks{Using data from the ESO VLT, programs 087.B-0439(A), 85.B-0615(A), and 71.B-0121(A).}}
\author[M. A. DiPompeo, M. S. Brotherton, C. De Breuck]{M. A. DiPompeo$^{1}$\thanks{e-mail: mdipompe@uwyo.edu (MAD); mbrother@uwyo.edu (MSB); cdebreuc@eso.org (CDB)}, M. S. Brotherton$^{1}$\footnotemark[2], C. De Breuck$^{2}$\footnotemark[2],  \\
$^{1}$ University of Wyoming, Dept. of Physics and Astronomy 3905, 1000 E. University, Laramie, WY 82071 \\
$^{2}$ European Southern Observatory, Karl Schwarzschild Strasse 2, 85748 Garching bei M\"{u}nchen, Germany}

\begin{document}

\date{Accepted ?; Recieved ?; in original form 2012 May}

\pagerange{\pageref{firstpage}--\pageref{lastpage}} \pubyear{2012}

\maketitle

\label{firstpage}

\begin{abstract}
We present new spectropolarimetric observations of 8 radio-loud broad absorption line (BAL) quasars, and combine these new data with our previous spectropolarimetric atlases (of both radio-loud and radio-quiet objects) in order to investigate the polarization properties of BAL quasars as a group.  The total (radio-selected) sample includes 36 (26) high-ionization and 22 (15) low-ionization BAL quasars.  On average, we confirm that broad emission lines are polarized at a level similar to or less than the continuum and broad absorption troughs are more highly polarized, but we note that these properties are not true for all individual objects.  Of the whole sample, 18 (31\%) have high ($>$2\%) continuum polarization, including 45\% of the LoBALs and 22\% of HiBALs.  We identify a few correlations between polarization and other quasar properties, as well as some interesting non-correlations.  In particular, continuum polarization does not correlate with radio spectral index, which suggests that the polarization is not due to a standard geometry and preferred viewing angle to BAL quasars.  The polarization also does not correlate with the amount of intrinsic dust reddening, indicating that the polarization is not solely due to direct light attenuation either.  Polarization does appear to depend on the minimum BAL outflow velocity, confirming the results of previous studies, and it may correlate with the maximum outflow velocity.  We also find that continuum polarization anti-correlates with the polarization in the \CIV\ broad emission and broad absorption.  These results suggest that the polarization of BAL quasars cannot be described by one simple model, and that the scatterer location and geometry can vary significantly from object to object.
\end{abstract}


\section{INTRODUCTION}
Optical and ultraviolet linear polarization, caused by asymmetric scattering, is potentially a powerful tool in constraining the structure and geometry of unresolved sources.  In particular it is often looked to in answering the questions surrounding the geometry of the approximately 20\% of optically-selected quasars with broad absorption lines (BALs; Knigge et al. 2008).  BAL quasars are an important subclass, because they represent those quasars with the most massive, highest velocity outflows.  Much recent work has shown that quasar outflows likely have effects on the evolution of the host galaxy, surrounding intergalactic medium, and quasar itself (Silk \& Rees 1998, King 2003, Hopkins \& Elvis 2010, Cano-Diaz et al. 2012, and others).  However, despite a few decades of study, we still do not understand the fundamental reason why only a fraction of quasars show BALs.

The two most popular explanations have involved either pure orientation/geometric effects, or pure evolutionary effects.  The orientation explanation has drawn heavily on the unification of type 1 and 2 Seyfert galaxies via viewing angle (Antonucci et al. 1993).  In this scenario, all quasars have BAL outflows, but only those seen from a more equatorial viewing angle are seen through the outflowing wind (e.g. Elvis 2000).  Equatorial is used somewhat loosely here, since most quasars from a true edge-on perspective are probably obscured by a dusty ``torus.''  We simply mean that they are seen from a viewing angle farther from the accretion disk symmetry axis compared to unabsorbed quasars.  

Previous polarization studies often argue in favor of this picture (e.g Ogle et al. 1999), particularly in light of the fact that BAL quasars are more often highly polarized than their non-BAL counterparts.  If a Seyfert-like scattering geometry is assumed, with scattering material located in a polar region, we would expect only the edge-on viewed sources (BALs in this scenario) to be highly polarized.  The analysis of Lamy \& Hutsemekers (2004), led them to a similar conclusion, developing a slightly more complex ``two-component wind'' model, with a less dense wind/outflow with more free electrons in a polar direction as the cause of the scattering.  

Weymann et al. (1991) also argued that their comparison of the emission-line properties of BAL and non-BAL quasars favored this scenario, as relatively small differences were found.  The differences discovered were that BAL quasars are more reddened, and they exhibit stronger \FeII\ and \AlIII\ emission; these results have been seen in radio-selected and strictly radio-loud BAL quasars as well (Brotherton et al. 2001, DiPompeo et al. 2012b).

While all of these studies are consistent with an orientation-only explanation, they do not necessarily require it.  In fact, particularly regarding observations at radio wavelengths, orientation-only scenarios have a multitude of of problems.  Radio-loud BAL quasars were only discovered slightly more than a decade ago (for example, Becker et al. 1997, Brotherton et al. 1998).  At least at moderate radio resolutions (such as the FIRST survey at 5\arcsec, Becker et al. 1995) they are more often spatially unresolved than non-BAL quasars (Becker et al. 2000).  Their relative rarity and compactness are difficult to reconcile with an equatorial line of sight.  Because of this, time-based explanations began to emerge, where the BAL phase is an early evolutionary phase of all quasar lifetimes.  In this picture, BAL quasars are young or rejuvenated quasars that are heavily enshrouded in a cocoon of gas and dust (Boroson \& Meyers 1992, Gregg et all 2002, 2006).  As radio jets develop and clear out this material, there is only a short overlap between the BAL and radio-loud quasar phase.

Aside from radio morphology and other properties such as radio core dominance, the shape of the radio spectrum (described by radio spectral index $\alpha_{rad}$; $S \propto \nu^{\alpha_{rad}}$, where $S$ is the radio flux and $\nu$ is the frequency) is a useful orientation indicator in unresolved sources.  Because of relativistic beaming effects, radio cores dominate the radio flux at small jet viewing angles (more pole-on), while radio jets dominate at larger viewing angles.  Radio cores tend to be optically thick and thus have flatter radio spectra ($\alpha_{rad} > -0.5$), while radio jets are optically thin with steep radio spectra ($\alpha_{rad} < -0.5$).  While there is significant scatter in the $\alpha_{rad}$-viewing angle relationship (DiPompeo et al. 2012a), spectral index is a useful diagnostic of ensemble orientations.

Until recently, no difference between BAL and non-BAL spectral index distributions had been found, suggesting that they were seen from similar ranges of viewing angles (Becker et al. 2000, Montenegro-Montes et al. 2008, Fine et al. 2011).  However, DiPompeo et al. (2011b) observed a sample specifically selected for this test with nearly twice as many objects and did find an overabundance of steep spectrum BAL quasars compared to a well-matched sample of non-BAL sources.  This suggests that at the largest viewing angles (before the quasar becomes completely obscured) one will generally see a BAL quasar.  However, the range of spectral indices for both samples was similar, indicating that BAL quasars were indeed sometimes seen along the same lines of sight normal quasars are seen from, including pole-on.  This was confirmed with modeling by DiPompeo et al. (2012a).  This result supports the claims of the existence of ``polar'' BALs identified using short timescale radio variability (Ghosh \& Punsly 2007, Zhou et al. 2006).  

Of course, all of this analysis relies on the assumption that the radio jets are perpendicular to the accretion disk plane.  However, there is evidence that this is indeed the case for type 2 AGN (di Serego Alighieri et al. 1993, Vernet et al. 2001), and is likely the case for higher redshift quasars (Drouart et al. 2012, submitted).

In addition to the above findings, other recent results have been pointing to a picture in which both orientation and evolution might play a role, but neither is sufficient on its own.  For example, Gallagher et al. (2007) found that BAL and unabsorbed quasars do not differ significantly in their infrared properties, suggesting that BAL sources are not significantly more enshrouded in dust.  Richards et al. (2011) argue that the direction and opening angle of BAL winds has an orientation dependence, but this is in turn driven by changes in the quasar SED, particularly at x-ray energies, which is a time-dependent phenomenon.  Allen et al. (2011) seem to have identified a redshift dependence to the BAL fraction, difficult to explain using only orientation.  Thus, it seems that all recent results are pointing toward a unification that depends on both orientation \textit{and} time.

In DiPompeo et al. (2010, 2011a) we presented spectropolarimetry of nearly 50 BAL quasars, but as separate samples.  Since that time, we have obtained new spectropolarimetric data of eight more sources; here we present this new data, and an in-depth analysis of the polarization properties of all of these objects as a group in order to constrain models describing the source of the polarization in BAL quasars.

We adopt the cosmology of Spergel et al. (2007) for all calculated properties, with $H_0 = 71$ km s$^{-1}$ Mpc$^{-1}$, $\Omega_M=0.27$ and $\Omega_{\Lambda} =0.73$.

\section{THE SAMPLE}
Because this analysis utilizes samples from several DiPompeo et al. references, we give them names as follows to help the reader keep track.  The sample of DiPompeo et al. (2010) is referred to as the ``Keck sample", the sample of DiPompeo et al. (2011a) is referred to as the ``VLT sample", the sample of DiPompeo et al. (2011b) is called the ``VLA sample", and the objects from the VLA sample with spectropolarimetry are referred to as the ``VLA-pol sample".  We provide below a brief summary of each.

The Keck sample contains 30 radio-selected BAL quasars.  The majority of these (27) are from the samples of Becker et al. (2000, 2001), which contains BAL quasars in the FIRST Bright Quasar Survey (FBQS; White et al. 2000).  Also included in this sample are 3 other radio-bright BAL quasars from the literature.  See DiPompeo et al. (2010) for more details.

The VLT sample contains 19 BAL quasars, as well as 4 sources originally considered to be possible BAL quasars.  As noted in DiPompeo et al. (2011a), they are more likely non-BAL quasars given the most recent observations, and so we will not include those sources in this analysis.  Three of the included objects are radio-selected.  It also contains some extreme objects, several found in Hall et al. (2002), which are particularly useful in extending the reddening parameter space investigated here.  Many objects are simply bright, southern BAL quasars which could be observed with the VLT in reasonable amounts of time.  We note that this sample is heterogeneous, but it is utilized because it significantly increases the sample size and parameter space covered; however, we will also discuss the results of the following analysis with this sample excluded due to its heterogeneous nature.  See DiPompeo et al. (2011a) for more details.

Presented along with the VLT sample was spectropolarimetry of two BAL quasars from the VLA sample.  We present here eight additional sources from that observing program.  We refer to DiPompeo et al (2011b) for details regarding the VLA sample; the VLA-pol sample includes the sources from there which are far enough south to observe at the VLT (lower than 10$^{\circ}$ declination).

Finally, we include FIRST source FIRST J155633.8+351758 in this analysis, which is in the VLA sample and has spectropolarimetry presented by Brotherton et al. (1997).  1556+3517 is one of the most highly polarized and reddened BAL quasars discovered to date (Becker et al. 1997).  Despite this, it still matches the selection criteria for the VLA sample.

We present basic properties of all sources in Table~\ref{proptbl}, with samples from different locations separated by horizontal lines.  The first section is the VLA-pol sample, including those with new observations presented here, the middle section is the Keck sample, and the last section is the VLT sample.  Note that there is some overlap between the Keck and VLA-pol samples; these objects are only presented once in the table and footnoted in the last column.  The columns are as follows: the source RA and DEC in columns 1 and 2, the redshift in column 3, the FIRST peak and integrated fluxes in columns 4-5, the absolute $B$ magnitude $M_B$ in column 6, the FIRST integrated flux k-corrected to a rest-frame 5 GHz radio luminosity in column 7, the radio loudness parameter $R^*$ (as defined by Stocke et al. 1992) in column 8, the radio spectral index $\alpha_{rad}$ in column 9, the balnicity index ($BI$) in column 10, the absorption index ($AI$) in column 11, the maximum velocity of the BAL outflow ($v_{max}$) in column 12, the minimum outflow velocity ($v_{min}$) in column 13, and the BAL subclass in column 10.  See section 3.3 for more information on $BI$, $AI$, $v_{max}$, and $v_{min}$.  

The classifications are as follows: HiBALs have evidence for BAL troughs from highly ionized species (in particular \CIV\ $\lambda$1549 \AA, but possibly also \SiIV\ $\lambda$1396 \AA\ and \NV\ $\lambda$1240 \AA), LoBALs have additional absorption from either \MgII\ $\lambda$ 2799 \AA\ and/or \AlIII\ $\lambda$ 1857 \AA, and FeLoBALs have absorption from various levels of \FeII.  Note that the values of $L_r$ for the Keck sample are slightly different than in the original reference; this is due to a small systematic correction to the luminosity distance.  The change in no way affects the correlation tests performed in DiPompeo et al. (2010) due to its systematic nature.

Some care has been taken when calculating $M_B$ and $R^{*}$ (which depends on the $B$ magnitude), as this bandpass often covers areas of heavy BAL absorption and reddening.  To account for this, we follow a similar method as Gregg et al. (2006).  We take average values of $g-i$ for quasars as a function of redshift from Richards et al. (2003) and use the $i$-band magnitude (which is much less effected by absorption and reddening) to calculate an unrededdened $g$ magnitude.  In cases where sources are not found in SDSS, $g$ and $i$ magnitudes are integrated from our total light spectra (all new sources presented here are from SDSS; 7 sources from the VLT sample are not, along with 3 from the Keck sample).  Both the Keck and VLT sample data have wavelength coverage that completely overlap both $g$ and $i$ bandpasses; no attempt to correct for slit losses is made, but this should not be very problematic at the red end of the spectra where we are making measurements.  If this new $g$-band magnitude is lower than the original (indicating that the source was likely reddened and/or significantly affected by LoBAL absorption), the new value is adopted.  Otherwise, we keep the original value to avoid making the objects bluer without a compelling reason to do so.  We then convert this to a $B$ band magnitude with the relation $B=g+0.47(g-r)+0.17$ (Smith et al. 2002), and use this to calculate $M_B$ and $R^{*}$.  More discussion of this can be found in section 3.3.

\setcounter{table}{0}
\begin{table*}\scriptsize
 \centering
  \caption{The full sample and basic properties.}
  \label{proptbl}
  \begin{tabular}{cccccccccccccccc}
  \hline
  RA             & DEC            & $z$  & $S_p$ & $S_i$ & $M_B$$^5$ & $\log L_r$ & $\log R^{*}$$^5$ & $\alpha_{rad}$ & $BI$ & $AI$ & $v_{max}$ & $v_{min}$ &   Class  \\
   (1)            &   (2)              &   (3)  &   (4)      &    (5)    &  (6)        &    (7)           &    (8)            &     (9)                                &   (10)     &  (11) &  (12) &       (13)       &   (14)         \\
 \hline
09 05 52 &  02  59 31 & 1.83 &  43.54 &  43.84 & -28.8 & 33.5 & 1.55 & -0.55 &   256 &   439 &  23807 & 11620 & HiBAL \\
10 44 52 &  10  40 05 & 1.88 &  16.40 &  17.21 & -28.0 & 33.1 & 1.45 & -0.92 &  3062 &  5570 &  22608 &   720 & HiBAL \\
11 02 06 &  11  21 04 & 2.35 &  82.32 &  83.06 & -27.8 & 34.0 & 2.41 & -1.32 &     0 &   371$^4$ &   1421 &  -785 & HiBAL \\
11 59 44 &  01  12 06 & 2.00 & 266.53 & 268.48 & -28.6 & 34.4 & 2.56 &  0.31 &     0 &  1715 &   4146 &   749 & HiBAL \\
12 28 48 & -01  04 14 & 2.66 &  29.36 &  30.81 & -28.4 & 33.7 & 1.90 & -0.45 &     4 &    12 &  8198 &  5638 & HiBAL \\
12 35 11 &  07  33 30 & 3.03 &  10.29 &  11.28 & -28.0 & 33.5 & 1.79 & -1.97 &     0 &  1235$^4$ &   3584 &  1307 & HiBAL \\
13 02 08 &  00  37 31 & 1.68 &  11.24 &  11.93 & -27.4 & 32.8 & 1.38 & -1.15 &   608 &  1163 &  9335 &  4353 & LoBAL \\
13 07 56 &  04  22 15 & 3.03 &  14.29 &  14.90 & -29.0 & 33.5 & 1.46 & -0.55 &   335 &   921 &  20840 & 14845 & HiBAL \\
13 37 01 &  02  46 30 & 3.06 &  43.32 &  44.82 & -28.0 & 34.1 & 2.40 & -1.76 &     0 &   732$^4$ &   4411 &   640 & HiBAL$^1$ \\
14 13 34 &  42  12 01 & 2.82 &  17.79 &  18.74 & -28.1 & 33.5 & 1.81 &  0.77 &     0 &  2212 &   3554 &   263 & HiBAL$^2$ \\
15 16 30 & -00  56 25 & 1.92 &  24.58 &  25.45 & -27.0 & 33.3 & 2.04 & -0.62 &     0 &   646$^4$ &   1842 &  -995 & HiBAL$^1$ \\
15 56 33 &  35  17 57 & 1.50 &  30.92 &  31.72 & -27.4 & 33.2 & 1.78 & -0.32 &   ... &   ... &    ... &   ... & FeLoBAL$^3$ \\
16 03 54 &  30  02 08 & 2.03 &  53.70 &  54.17 & -27.8 & 33.7 & 2.14 & -0.57 &     0 &  1105 &   3564 &   907 & HiBAL$^2$ \\
16 55 43 &  39  45 19 & 1.75 &  10.15 &  10.16 & -27.3 & 32.9 & 1.51 & -0.20 &  4134 &  4784 &  22874 &  9213 & HiBAL$^2$ \\
\hline
01 35 15 & -02  13 49 & 1.82 &  22.42 &  22.81 & -28.0 & 33.3 & 1.62 &  0.18 &  2140 &  2506 &  13838 &   800 & HiBAL \\
02 56 25 & -01  19 04 & 2.49 &  25.78 &  27.56 & -27.8 & 33.6 & 2.02 & -0.75 &   328 &   725 &  12837 &  1000 & HiBAL \\
07 24 17 &  41  59 18 & 1.55 &   7.89 &   7.90 & -27.0 & 32.7 & 1.44 &  0.00 &  $>$5061 & $>$5596 & $>$18950 &  8980 & LoBAL \\
07 28 30 &  40  26 22 & 0.66 &  16.96 &  16.79 & -27.9 & 31.9 & 0.28 & -1.10 &   ... &   ... &    ... &   ... & LoBAL \\
08 09 01 &  27  53 54 & 1.51 &   1.17 &   1.67 & -27.4 & 31.9 & 0.47 & ... &  $>$298 & $>$1017 & $>$13724 & 10200 & HiBAL \\
09 10 44 &  26  13 01 & 2.94 &   7.84 &   7.46 & -27.6 & 32.7 & 1.23 & -0.50 &    66 &  2414 &   5158 &   500 & HiBAL \\
09 13 29 &  39  44 42 & 1.58 &   2.06 &   2.09 & -27.4 & 32.0 & 0.61 & -0.60 & $>$5984 & $>$6362 & $>$21951 &  2900 & HiBAL \\
09 34 04 &  31  53 42 & 2.42 &   4.68 &   4.39 & -28.8 & 32.8 & 0.81 & -0.20 &  1092 &  1806 &  25000 &  8600 & HiBAL \\
09 46 02 &  27  44 03 & 1.74 &   3.54 &   3.63 & -28.1 & 32.4 & 0.67 & ... &     0 &   197$^4$ &  6897 &   700 & HiBAL \\
09 57 07 &  23  56 20 & 1.99 & 136.10 & 140.49 & -27.9 & 34.1 & 2.46 & -0.60 &     0 &  2884 &   3834 &  -280 & HiBAL \\
10 16 12 &  52  09 22 & 2.46 &   5.26 & 176.84 & -26.7 & 34.4 & 3.23 & -1.00 &  2439 &  5344 &  16466 & -1290 & HiBAL \\
10 31 10 &  39  53 21 & 1.08 &   2.45 &   2.03 & -25.8 & 31.7 & 0.97 & -0.20 &   ... &   ... &    ... &   ... & LoBAL \\
10 44 59 &  36  56 00 & 0.70 &  14.61 &  15.00 & -26.3 & 32.1 & 1.15 & -0.50 &   ... &   ... &    ... &   ... & FeLoBAL \\
10 54 27 &  25  36 23 & 2.40 &   2.99 &   3.02 & -28.1 & 32.6 & 0.89 & -0.50 &  2531 &  6222 &   24980 &  1500 & LoBAL \\
11 22 20 &  31  24 57 & 1.45 &  12.64 &  12.87 & -27.2 & 32.7 & 1.39 & -0.60 &     0 & $>$1068 &  $>$6033 &   145 & LoBAL \\
12 00 51 &  35  08 36 & 1.70 &   2.03 &   1.46 & -28.6 & 31.9 & 0.04 & -0.80 &  2845 &  4683 &  12289 &  1750 & HiBAL \\
12 14 42 &  28  03 44 & 0.70 &   2.61 &   2.90 & -25.9 & 31.3 & 0.51 & -0.80 &   ... &   ... &    ... &   ... & FeLoBAL \\
13 12 13 &  23  20 25 & 1.51 &  43.27 &  44.12 & -27.8 & 33.3 & 1.73 & -0.80 &     0 &   $>$15 & $>$14107 & 10850 & HiBAL \\
14 08 00 &  34  51 25 & 1.22 &   2.91 &   2.87 & -27.0 & 31.9 & 0.66 & -0.60 &   ... &   ... &    ... &   ... & LoBAL \\
14 08 07 &  30  54 39 & 0.84 &   3.34 &   3.21 & -26.1 & 31.6 & 0.68 & -0.70 &   ... &   ... &    ... &   ... & LoBAL \\
14 20 14 &  25  33 54 & 2.20 &   1.25 &   1.20 & -28.4 & 32.1 & 0.28 & -1.10 &  2860 &  3635 &  25000 &   350 & HiBAL \\
14 27 03 &  27  09 40 & 1.17 &   2.58 &   2.98 & -26.0 & 31.9 & 1.01 & -0.70 &   ... &   ... &    ... &   ... & FeLoBAL \\
15 23 15 &  37  59 20 & 1.34 &   1.67 &   1.83 & -26.9 & 31.8 & 0.61 & -0.60 &   ... &   ... &    ... &   ... & LoBAL \\
15 23 50 &  39  14 04 & 0.66 &   3.75 &   4.07 & -26.3 & 31.5 & 0.55 & -0.40 &   ... &   ... &    ... &   ... & LoBAL \\
16 41 52 &  30  58 51 & 2.00 &   2.13 &   2.66 & -27.6 & 32.5 & 0.95 &  0.50 &  7868 & 12082 &  22631 &  -220 & LoBAL \\
17 09 19 &  28  18 35 & 2.37 &   1.51 &   2.15 & -28.0 & 32.4 & 0.77 & ... &   207 &  2079 &   5590 &   900 & HiBAL \\
23 59 52 & -12  41 37 & 0.87 & ... & ... & -27.0 & ... & ... & -0.36 &   ... &   ... &    ... &   ... & FeLoBAL \\
\hline
00 01 21 &  02  33 05 & 1.87 & ... & ... & -27.0 & ... & ... & ... &  5892 &  6806 &  19301 &  3700 & LoBAL \\
00 18 24 &  00  15 26 & 2.43 & ... & ... & -26.5 & ... & ... & ... &  5189 &  7633 &  18806 &   700 & HiBAL \\
00 43 23 & -00  15 52 & 2.80 & 130.68 & 158.10 & -28.1 & 34.5 & 2.76 & ... &  1006 &  1204 &  21566 &  6800 & HiBAL \\
00 49 38 & -30  39 51 & 2.36 & ... & ... & -27.1 & ... & ... & ... &  5681 &  8384 &  17828 &   600 & HiBAL \\
01 12 27 & -01  12 20 & 1.76 & ... & ... & -27.3 & ... & ... & ... &  2877 &  3955 &  24877 &  2800 & HiBAL \\
01 15 39 & -00  33 27 & 1.59 & ... & ... & -27.2 & ... & ... & ... &  2737 &  3173 &  20873 &  8900 & HiBAL \\
01 36 52 & -37  25 04 & 2.52 & ... & ... & -28.8 & ... & ... & ... & 10028 & 11219 &  25000 &  5300 & HiBAL \\
02 50 42 &  00  35 38 & 2.38 & ... & ... & -27.4 & ... & ... & ... &  4275 &  7358 &  14573 &  -125 & LoBAL \\
03 00 00 &  00  48 33 & 0.89 & ... & ... & -27.1 & ... & ... & ... &   ... &   ... &    ... &   ... & FeLoBAL \\
03 18 56 & -06  00 36 & 1.97 & ... & ... & -28.4 & ... & ... & ... &  4164 &  5662 &  11276 &  1900 & FeLoBAL \\
03 37 21 & -33  29 11 & 2.26 & ... & ... & -27.1 & ... & ... & ... &  7939 &  8567 &  24896 &   200 & HiBAL \\
20 46 44 & -34  33 35 & 3.35 & ... & ... & -27.5 & ... & ... & ... &  1898 &  6475 &  23968 &   300 & HiBAL \\
21 15 06 & -43  23 10 & 1.71 & ... & ... & -28.9 & ... & ... & ... &  3334 &  4614 &  18835 &  2200 & HiBAL \\
22 15 11 & -00  45 48 & 1.48 & ... & ... & -28.1 & ... & ... & ... &  7011 &  7209 &  21497 &  1800 & LoBAL \\
22 43 46 & -36  47 03 & 1.83 & ... & ... & -28.6 & ... & ... & ... &  7807 & 12148 &  24183 & -1054 & LoBAL \\
23 52 38 &  01  05 53 & 2.16 & ... & ... & -28.7 & ... & ... & ... &     0 &  1913 &   4095 &   450 & HiBAL \\
23 52 53 & -00  28 50 & 1.62 & ... & ... & -26.7 & ... & ... & ... &  8385 &  8906 &  25000 &  8140 & HiBAL \\
 \hline
 \end{tabular}
 \\
 {
 \raggedright
 $^1$The spectropolarimetry for these objects was presented along with the VLT sample. \\
 $^2$These sources happened to overlap with the objects in the Keck sample, and their spectropolarimetry information can be found in the original reference. \\
 $^3$The spectropolarimetry for this source was presented by Brotherton et al. (1997). \\
 $^4$$AI$ was integrated using a condition of 800 km/s for continuous absorption. \\
 $^5$Corrected for reddening/absorption using SDSS colors and $i$-band magnitudes- see sections 2 and 3.3. \\
  The horizontal lines split the table into three groups; the objects from the VLA-pol sample, including those with new observations presented here, the objects from the Keck sample, and the objects from the VLT sample, respectively.  $S_p$ and $S_i$ are the 1.4 GHz peak and integrated FIRST fluxes.  $L_r$ is the k-corrected 5GHz rest-frame radio luminosity, using the FIRST integrated flux measurement and the spectral index listed in column 9.  $R^*$ is the radio loudness parameter, as defined by Stocke et al. (1992).  $\alpha_{rad}$ is the radio spectral index.  See the text for a discussion of $BI$, $AI$, $v_{max}$, and $v_{min}$. \\
  }
 \end{table*}

\section{OBSERVATIONS \& MEASUREMENTS}
\subsection{Spectropolarimetry observations}
The new observations here were part of the same program as SDSS J1337-0246 and SDSS J1516-0056 from DiPompeo et al. (2011a), with the same instrumental configuration, but observed in the following period.  Observations were performed mostly in April 2011, with the exception of 1307+0422 which was observed in July 2011.  

All observations were done in service mode with the PMOS mode of the FORS2 spectrograph (Appenzeller et al. 1998) on the Antu unit of the VLT.  Total exposure times (sum of all waveplate positions) ranged from 9 to 50 minutes, depending on the source V-band magnitude.  We used a 600 line mm$^{-1}$ grism blazed at 5000 \AA\ with a 1.4\arcsec\ slit, giving a resolution of 10 \AA\ and a dispersion of 2 \AA\ per pixel.  We note that when using this grism in PMOS mode, there is an issue with reflected light in the instrument optics that affects the observed wavelengths 4050-4110 \AA.  This is removed as best as possible when flat-fielding, but no polarization measurements are made in this region.

The wavelength dependence of the polarization position angle due to the instrument optics was calibrated with an ESO-supplied rotation curve.  The polarized standards\footnote{See http://www.eso.org/sci/facilities/paranal/instruments/fors/inst/pola.html for standard star information.} Hiltner 652 and Vela 1 were used to check the polarization position angle offset and polarization percentage.  Continuum polarizations are accurate to about 0.2\% and the position angle is accurate to about 1$^{\circ}$.  The unpolarized standard star WD1620-391 was observed and was polarized less than 0.08\%. We refer the reader to DiPompeo et al. (2010, 2011a) for full observational details for the rest of the sample.

\subsection{Polarization measurements}
Here we summarize the polarization measurements made, and full details are found in DiPompeo et al. (2010, 2011b).  After reducing the data to one-dimensional spectra using standard techniques and IRAF\footnote{IRAF is distributed by the National Optical Astronomy Observatory, which is operated by the Association of Universities for Research in Astronomy, Inc., under cooperative agreement with the National Science Foundation.} packages, standard procedures for calculating Stokes Q and U parameters were used (Miller et al. 1988, Cohen 1997).  Binning in flux (counts) is performed before calculating Stokes parameters.  Polarizations presented are de-biased using rotated Q and U parameters, and errors are 1$\sigma$ confidence intervals (Simmons \& Stewart 1985).

\begin{figure*}
   \subfloat[][Fig. 1$a$]{\includegraphics[width=5in]{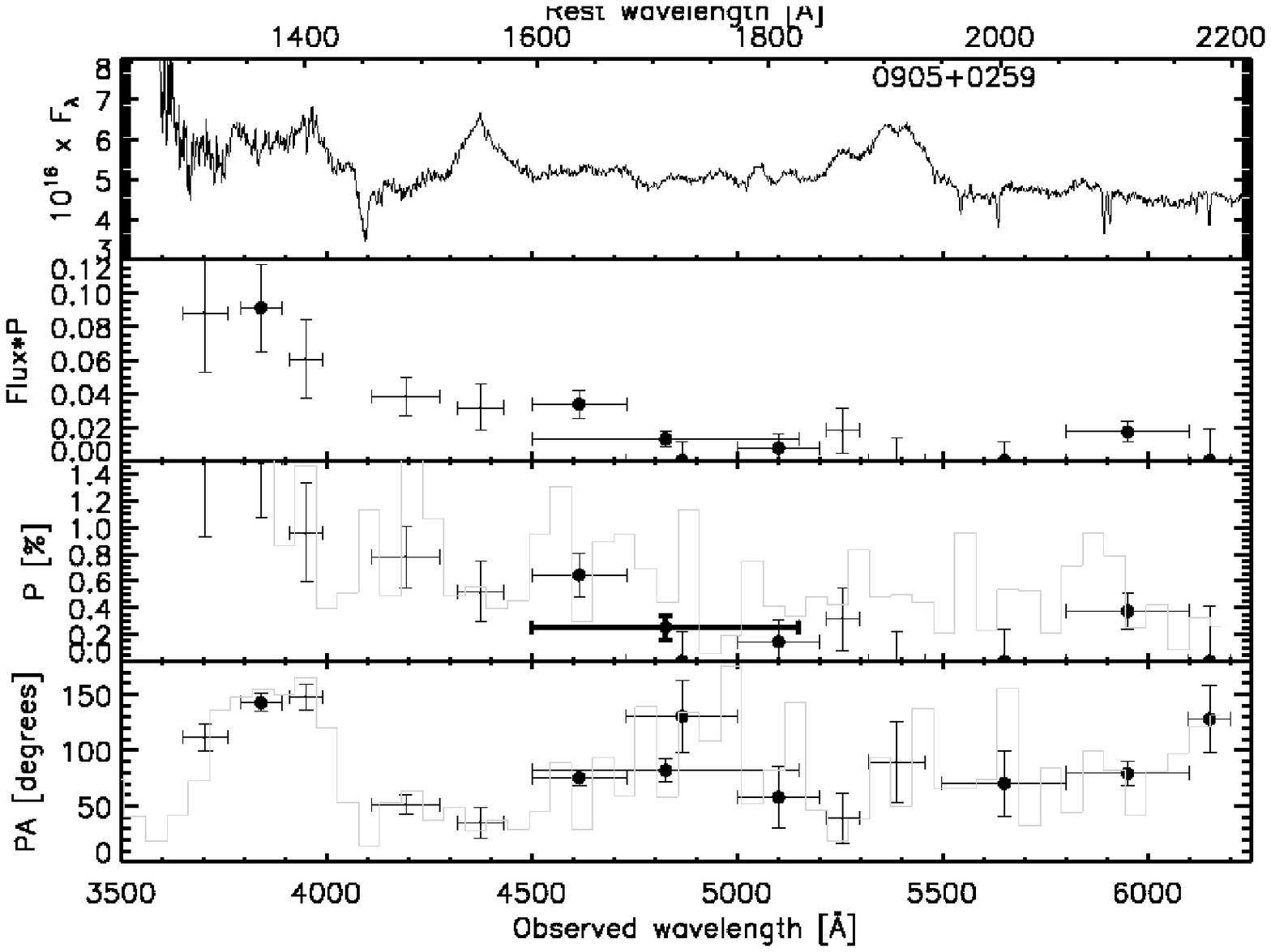}}
  \caption{The following panels show the the newly acquired spectropolarimetry, used in addition to the Keck and VLT samples.  In all figures, the top panel is the total flux spectrum (object names are in this panel), the polarized flux, the polarization percentage, and the polarization position angle.  The bold bins in the polarization percentage panel show where continuum polarization measurements are made.  The light grey lines in the last two panels are smaller binned, unbiased polarization measurements.\label{polplotsfig}}
  \label{polfig}
\end{figure*}

Data were binned when making the polarization measurements in order to reduce errors.  Continuum measurements were generally binned across 100-300\AA\, and emission and absorption features were binned as small as possible without getting errors too large to distinguish the measurements from surrounding spectral regions.  Figures~\ref{polfig} (a-h) show the spectropolarimetry results for the 8 new observations; see DiPompeo et al. (2010, 2011a) and Brotherton et al. (1997) for similar plots for the rest of the sample.  The panels in each figure are as follows (from top to bottom): the total flux spectrum (object names are included in the top corner of this panel), the polarized flux, the polarization percentage, and the polarization position angle.  The light grey lines in the polarization percentage and position angle panels show smaller bins (20-30 \AA) of unbiased measurements, which generally follow the larger bins closely but with larger errors (not shown).

Table~\ref{contpoltbl} lists the white light and continuum polarization properties of the newly observed objects.  White light measurements are averaged over the whole spectrum, from 3800-8500 \AA.  Continuum measurements are made over the indicated observed-frame wavelengths, and these bins are also shown in bold in the polarization percentage panels of Figure~\ref{polfig}.  The final column of the table gives the maximum theoretical interstellar polarization (ISP), based on Serkowsky et al. (1975) and $E(B-V)$ values from Schlegel et al. (1998) via NED\footnote{The NASA/IPAC Extragalactic Database (NED) is operated by the Jet Propulsion Laboratory, California Institute of Technology, under contract with the National Aeronautics and Space Administration}.  Table~\ref{linepoltbl} gives the polarization properties of the \CIV\ and \CIII\ emission lines ($p_e$) and the \CIV\ absorption trough ($p_a$).  These values are also given when normalized by the continuum immediately to the red side of the feature (unless only the blue side was available, in which case the measurement will have a footnote), in order to account for any wavelength dependence of the polarization when comparing objects.  These measurements are labeled as $p_e/p_c$ and $p_a/p_c$.  Note that the continuum in this normalization is not necessarily the same as the continuum polarization reported in Table~\ref{contpoltbl}.  Similar tables for the rest of the sample are found in DiPompeo et al. (2010, 2011a).  None of these new observations had \MgII\ available for measurement.

\vspace{3mm}
\noindent \textit{Notes on individual objects}
\vspace{3mm}

\noindent \textbf{SDSS J090552.40+025931.4} It is clear from Figure~\ref{polplotsfig} (a) and Table~\ref{contpoltbl} that SDSS J0905+0259 is likely intrinsically unpolarized, or only very weakly so.  The continuum polarization is at a similar level to the maximum possible ISP (the white light polarization is even lower), and the polarization position angle changes wildly across the spectrum.  Only on the far blue end of the spectrum does the polarization appear to rise to significant amounts, but it is not very convincing.

\begin{figure*}
  \subfloat[][Fig. 1$b$]{\includegraphics[width=4.5in]{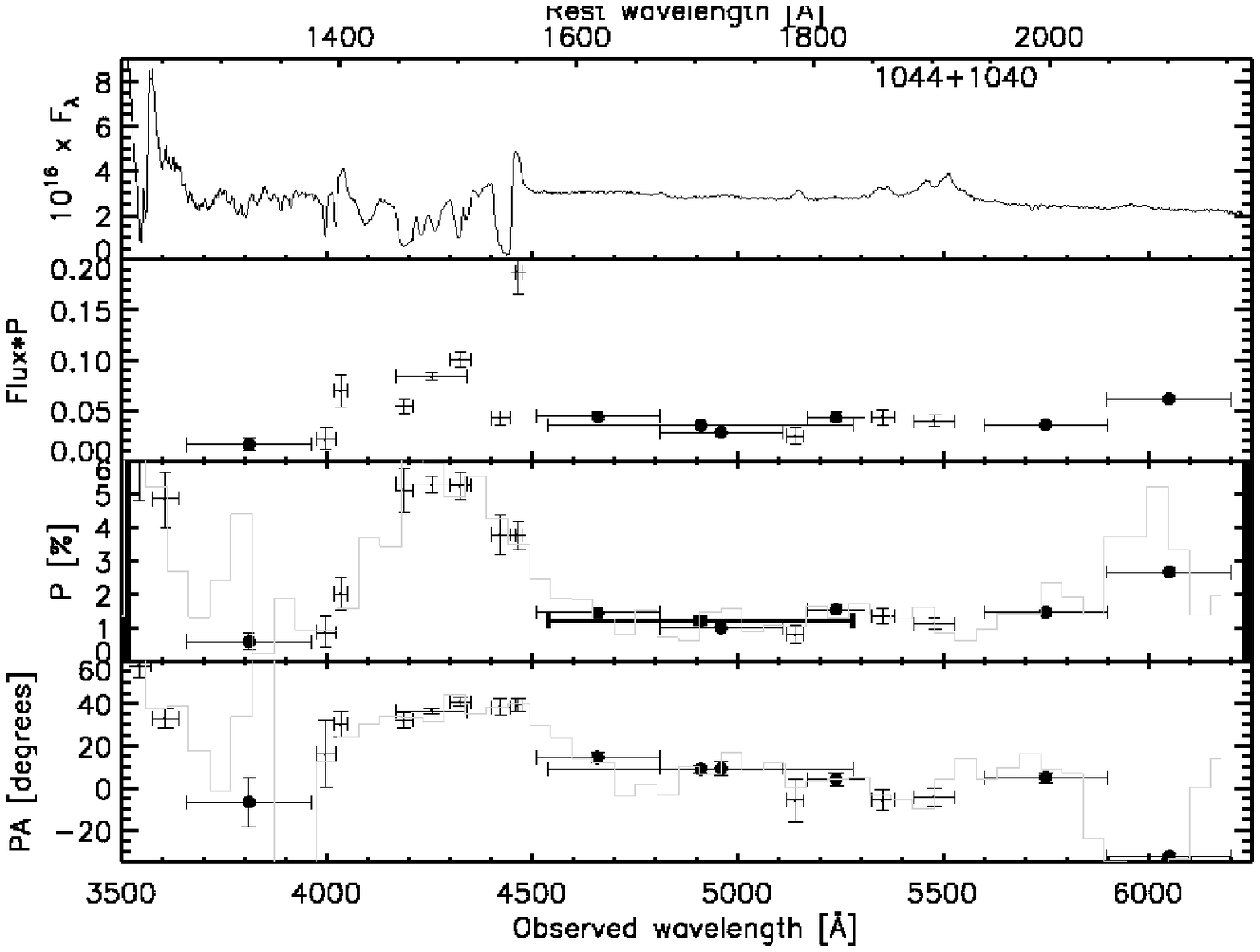}}
\end{figure*}

\begin{figure*}
  \subfloat[][Fig. 1$c$]{\includegraphics[width=4.5in]{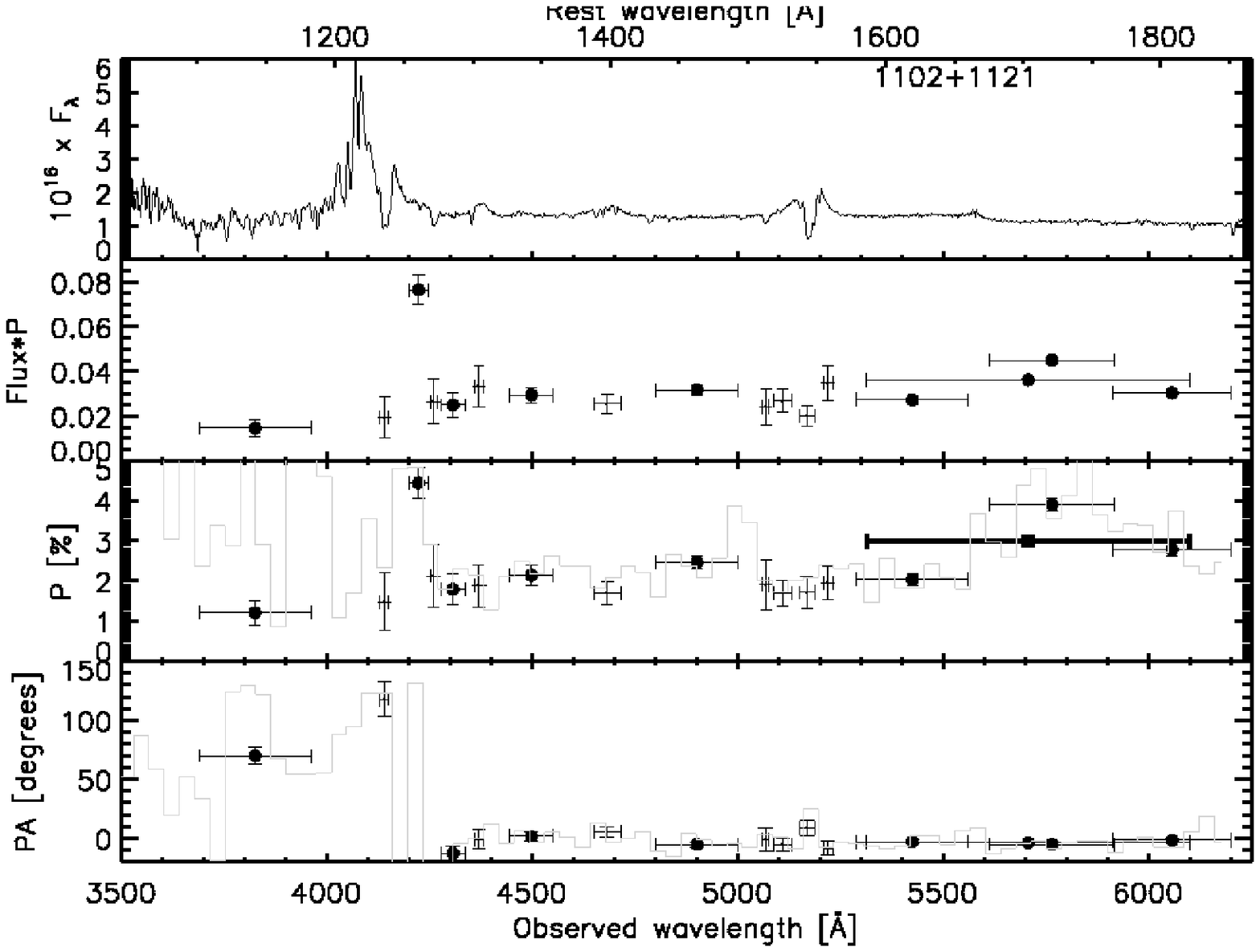}}
\end{figure*}

\begin{figure*}
  \subfloat[][Fig. 1$d$]{\includegraphics[width=4.5in]{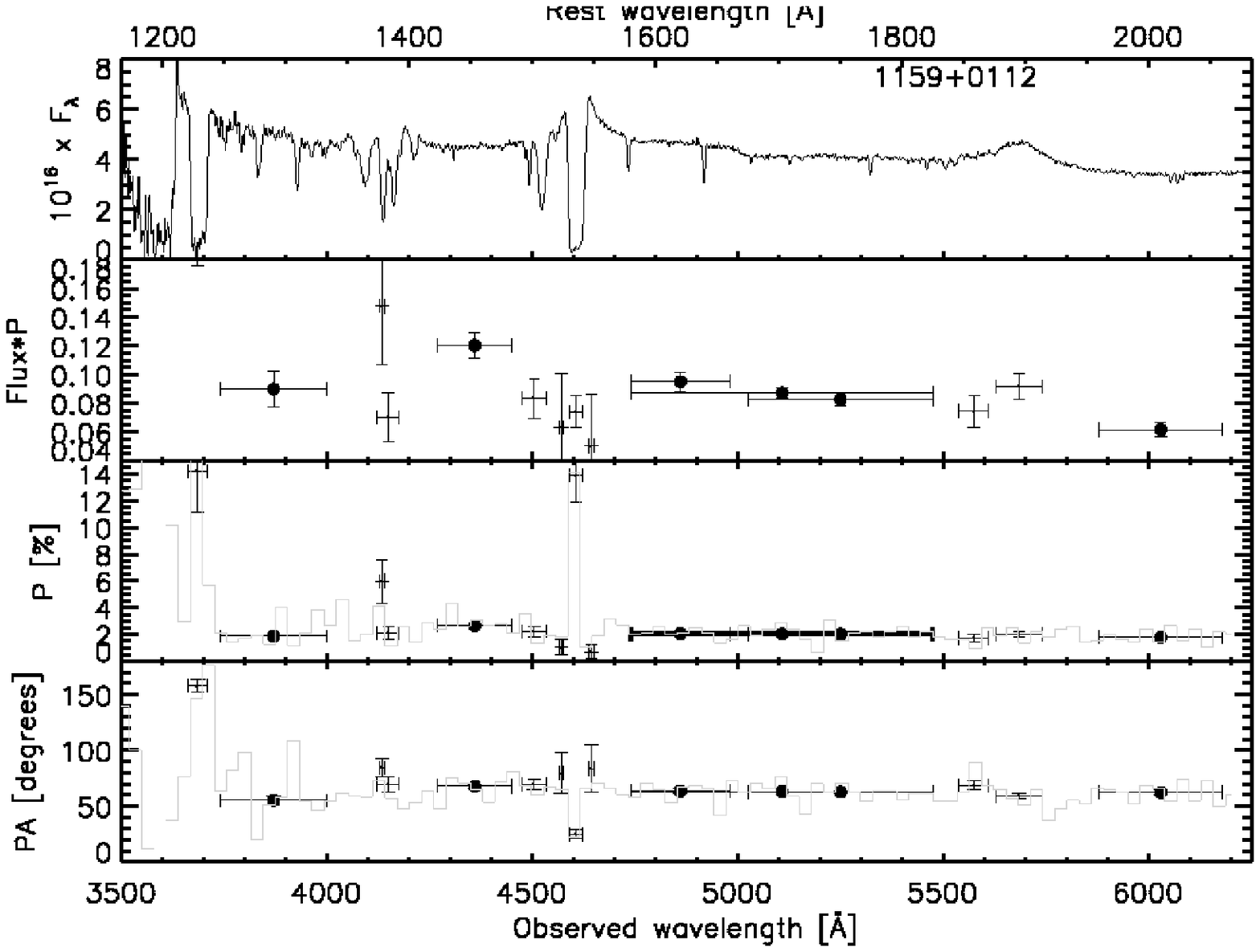}}
\end{figure*}

\begin{figure*}
  \subfloat[][Fig. 1$e$]{\includegraphics[width=4.5in]{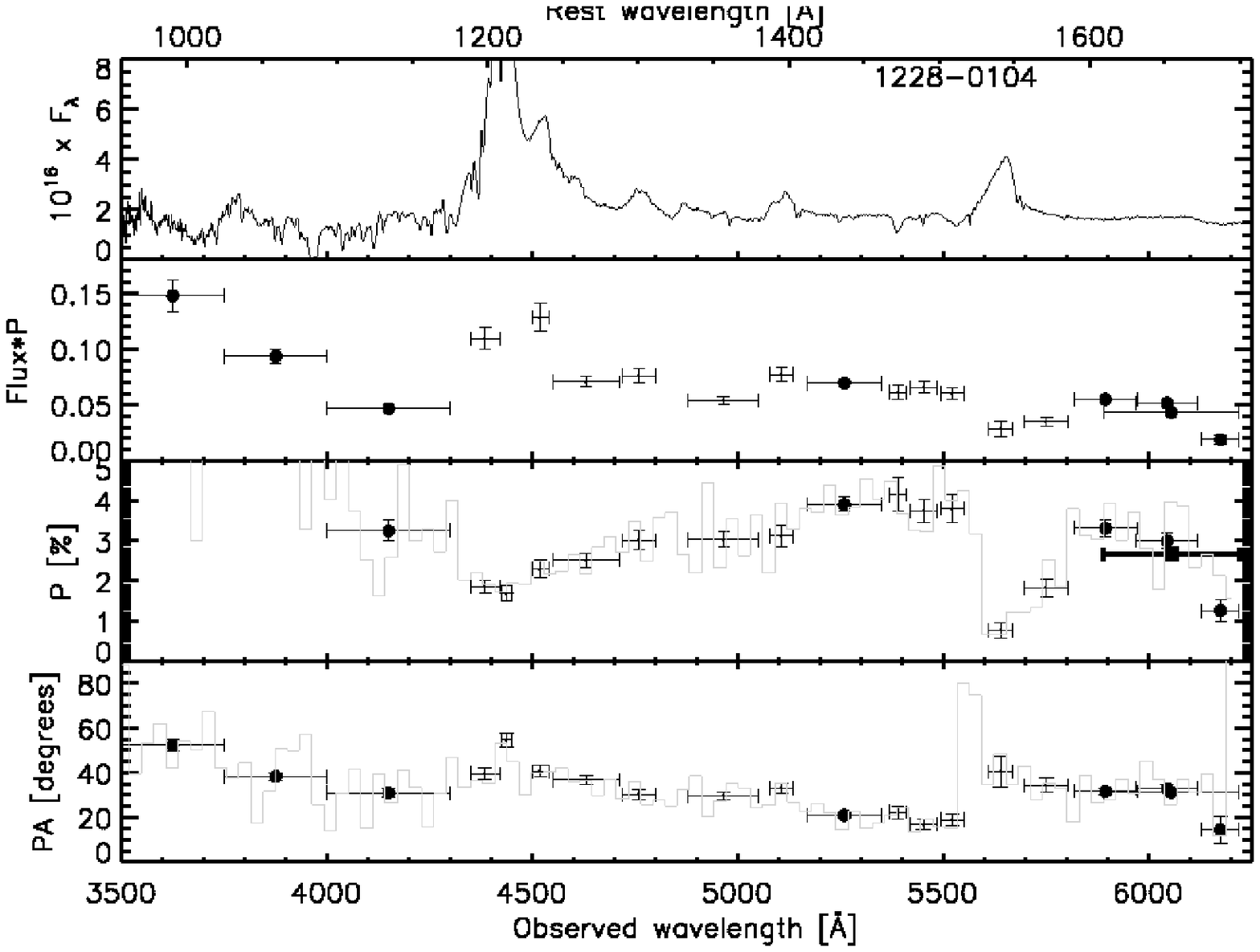}}
\end{figure*}

\begin{figure*}
  \subfloat[][Fig. 1$f$]{\includegraphics[width=4.5in]{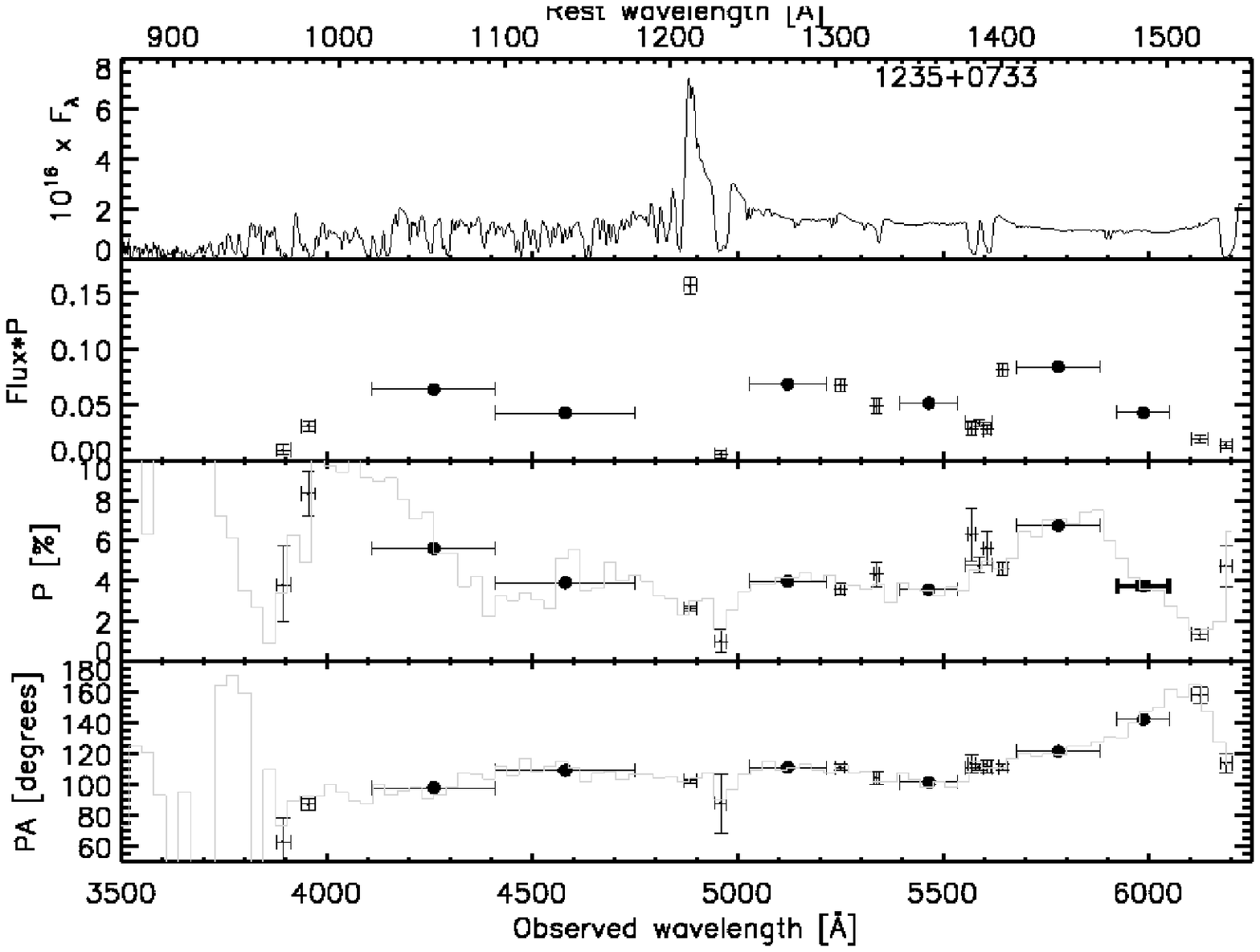}}
\end{figure*}

\begin{figure*}
  \subfloat[][Fig. 1$g$]{\includegraphics[width=4.5in]{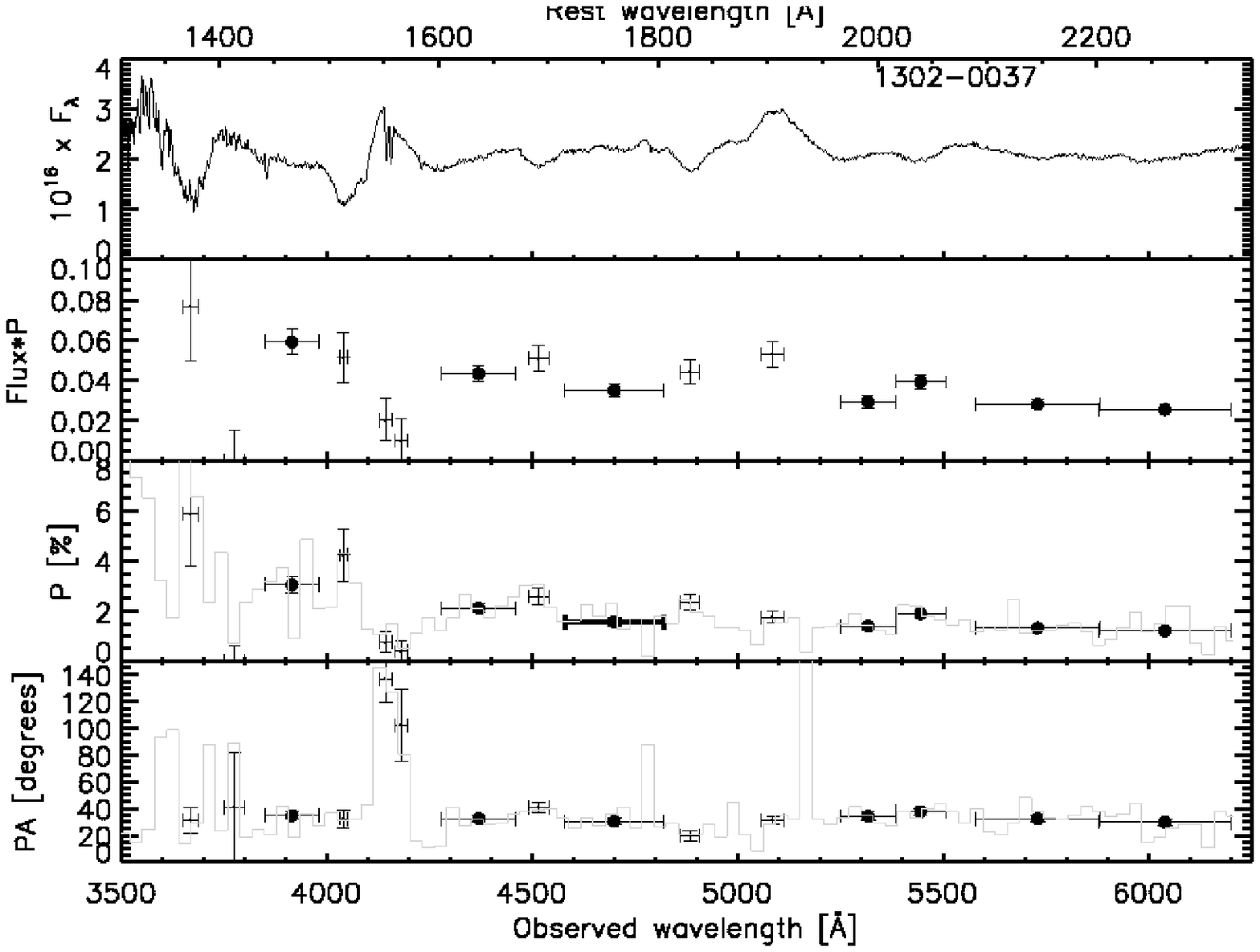}}
\end{figure*}

\begin{figure*}
  \subfloat[][Fig. 1$h$]{\includegraphics[width=4.5in]{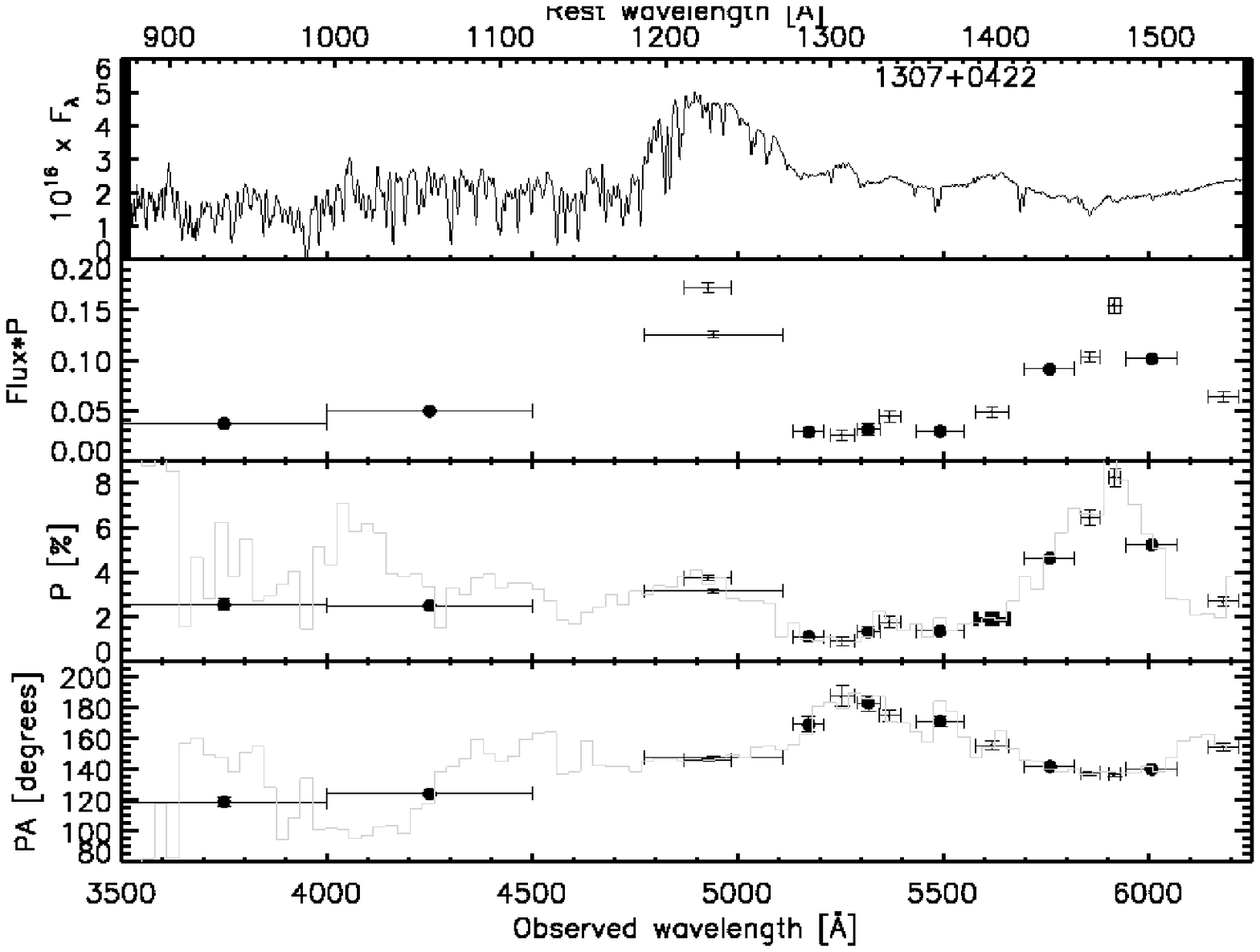}}
\end{figure*}

\setcounter{table}{1}
\begin{table*}
  \caption{Continuum and white light polarization properties of the new additions to sample.}
  \label{contpoltbl}
  \begin{tabular}{ccccccc}
  \hline
   Object & White P(\%) & White PA($^{\circ}$) & Cont. P(\%) & Cont. PA($^{\circ}$) & Cont. $\lambda$(\AA) & Max. ISP (\%) \\
  \hline
0905$+$0259 & 0.16$\pm$0.05 &  60$\pm$ 9  & 0.25$\pm$0.09 &  82$\pm$10 & 4500-5150 & 0.27 \\
1044$+$1040 & 1.21$\pm$0.04 &  10$\pm$ 1  & 1.22$\pm$0.07 &   9$\pm$ 2    & 4540-5280 & 0.25 \\
1102$+$1121 & 1.97$\pm$0.05 & 176$\pm$ 1 & 2.99$\pm$0.09 & 176$\pm$ 1 & 5315-6100 & 0.14 \\
1159$+$0112 & 2.00$\pm$0.05 &  62$\pm$ 1  & 2.04$\pm$0.08 &  63$\pm$ 1   & 4740-5475 & 0.19 \\
1228$-$0104  & 2.58$\pm$0.05 &  33$\pm$ 1  & 2.66$\pm$0.13 & 31$\pm$ 1    & 5890-6220 & 0.21 \\
1235$+$0733 & 3.69$\pm$0.03 & 111$\pm$ 0 & 3.58$\pm$0.16 & 142$\pm$ 1 & 5925-6050 & 0.22 \\
1302$-$0037  & 1.45$\pm$0.04 &  32$\pm$ 1  & 1.57$\pm$0.13 &  30$\pm$ 2   & 4580-4820 & 0.20 \\
1307$+$0422 & 2.51$\pm$0.04 & 146$\pm$ 1 & 1.36$\pm$0.17 & 171$\pm$ 4 & 5580-5660 & 0.23 \\
  \hline
  \end{tabular}
 \\
 {
 \raggedright
 White light polarization measurements are averaged over the whole spectrum, from 3800-8500 \AA.  Continuum polarization measurements are averaged over the observed wavelength region listed in column 6.  Maximum interstellar polarizations along the line of sight using the Serkowsky Law are listed in column 7. \\
 }
 \end{table*}

\setcounter{table}{2}
\begin{table*}
  \caption{Polarization properties of emission and absorption features in the new additions to the sample.}
  \label{linepoltbl}
  \begin{tabular}{ccccccccccc}
  \hline
    Object & $p_e$ (\CIV) & $p_e$ (\CIII) & $p_a$ (\CIV)  & $\frac{p_e}{p_c}$ (\CIV) & $\frac{p_e}{p_c}$ (\CIII) & $\frac{p_a}{p_c}$ (\CIV)  \\
   \hline
   0905$+$0259                                  & 0.51        & 0.00       &  ...         &  0.80      &   0.00        &  ...           \\
   1044$+$1040                                  & 3.77        & 1.12       & 5.29     &  2.59      &   0.76        &   3.64     \\
   1102$+$1121                                  & 1.83        &  ...           & 1.91     &  0.89      &   ...             &   0.84     \\
   1159$+$0112                                  & 0.87        & 2.00       & 13.91   &  0.87      &   1.12        &   6.78     \\
   1228$-$0104                                   & 0.77        & ...            & 4.14     & 0.29       &  ...              &  1.56      \\
   1235$+$0733$^{1}$                       & 1.37       &  ...            & 4.75     &   0.36     &  ...              &   1.26     \\
   1302$-$0037                                   & 0.38       & 1.77        & 4.23     &   0.18     &   1.27        &   2.00     \\
   1307$+$0422$^{1}$                       &  ...           &  ...            & 8.26     &   ...          &  ...              &   6.07     \\
   \hline
   \end{tabular}
   \\
   {
   \raggedright
     $^{1}$The continuum polarization for normalization in these objects was measured blueward of the emission lines because the lines are at the red edge of the spectrum.\\
  Typical uncertainties on $p_e$ and $p_a$ are approximately 0.2-0.4\%.  See text for discussion of all parameters.  No measurements of polarization in \MgII\ are measured here as they are not in the spectral window. \\
  }
\end{table*}

\subsection{Spectral measurements}
In addition to the polarization, we have made other measurements using the total flux spectra obtained during the observations.  All spectra are shifted to rest-frame wavelengths using the values of $z$ in Table~\ref{proptbl}.  In particular we measure several properties of the \CIV\ BAL troughs when it is present in the spectral window.  While the traditional measurements such as balnicity index ($BI$; Weymann et al. 1991) or absorption index ($AI$; e.g. Hall et al. 2002) are somewhat arbitrarily defined (Ganguly \& Brotherton 2008) and difficult to measure because they depend heavily on accurate placement of the continuum, we include them here for completeness.  $BI$ is calculated using:
\begin{equation}
 BI = \int_{3,000}^{25,000} \left(1-\frac{f(v)}{0.9} \right) C dv
\end{equation}
where $f(v)$ is the continuum normalized flux as a function of velocity (in km/s), and $C$ is equal to 1 when the normalized flux is below 0.9 continuously for 2000 km/s, otherwise it is set to 0.  The continuum is defined by eye, using a 3rd-5th order polynomial in spectral regions relatively free of strong emission or absorption features.  Note that $BI$ is a rather conservative measurement, and was originally developed to identify sources with very strong absorption in low resolution, low SNR spectra.  Many obvious BAL quasars can have a $BI$ of 0, particularly if the absorption begins at low velocity.  To account for this, $AI$ was developed:
\begin{equation}
 AI = \int_{0}^{25,000} \left(1-\frac{f(v)}{0.9} \right) C' dv
\end{equation}
Notice that the integration begins at 0 velocity here (but will still miss absorption that begins to the red of the emission peak).  For consistency with the measurements of others, we also use a requirement of continuous of absorption for 2000 km/s for $C'$ to be set to one, but we note that this is not necessarily part of a strict definition of $AI$.  There are a few situation in which this returns an $AI$ of 0, despite the clear presence of broad absorption.  This is often because much of the absorption is on top of the \CIV\ emission, and so the part of the absorption that drops below the continuum is slightly narrower than 2000 km/s.  We find that relaxing our continuous absorption requirement in these cases to 800 km/s allows us to obtain reasonable measures of $AI$ (and $v_{max}$; see below) without including any extra non-BAL absorption.  When this condition is used the measurements are footnoted in the table.  Both $AI$ and $BI$ are only integrated to 25,000 km/s because this is where the \SiIV\ complex begins.  

In addition, we measure the maximum and minimum outflow velocities ($v_{max}$ and $v_{min}$, respectively).  $v_{min}$ is measured in a somewhat subjective manner, in order to account for situations where absorption begins high on the \CIV\ emission line and would be missed using absorption only below the continuum.  Using the location where $C$ or $C'$ first become equal to 1 as the minimum can bias $v_{min}$ high due to this effect.  We measure $v_{max}$ using the location where $AI$ integration ends.  All of these measurements assume $v=0$ at 1549.06 \AA.  Table~\ref{proptbl} lists the values of $BI$, $AI$, $v_{max}$, and $v_{min}$.

We estimate the amount of intrinsic reddening in each source in two ways.  First, we use the difference between the original and unrededdened $g$-band magnitudes ($A_g$) that were discussed at the end of section 2.  We also estimate $A_V$ in all sources using the spectra.  First, all spectra are corrected for galactic reddening using the $E(B-V)$ values. We then utilize an SMC extinction curve (with $R_V=3.1$) and the composite spectrum of all quasars in the FBQS made available by Brotherton et al. (2001) to estimate $A_V$.  We chose this composite because the majority of our sources are radio-selected.  It is difficult to automate this procedure due to the heavy absorption in BAL quasar spectra, and so this was done by eye.  We simply stepped through values of $A_V$ in increments of 0.1 (after making an initial first guess) and de-reddened each spectrum, comparing each step with the FBQS composite until a reasonable agreement was found across the whole spectrum.  In some cases steps of 0.01 were used to fine-tune the reddening.  While it is impossible in most cases to get a perfect match to the composite, this method provides a reasonable estimate of the intrinsic reddening.  Our measures of $A_g$ and $A_V$ correlate strongly, indicating that these two independent methods provide robust estimates of the amount of reddening.  $A_g$ is however systematically larger, due to the fact that it is also correcting for flux lost due to BAL absorption in many cases.

\section{ANALYSIS \& RESULTS}
\begin{figure}
 \includegraphics[width=3.2in]{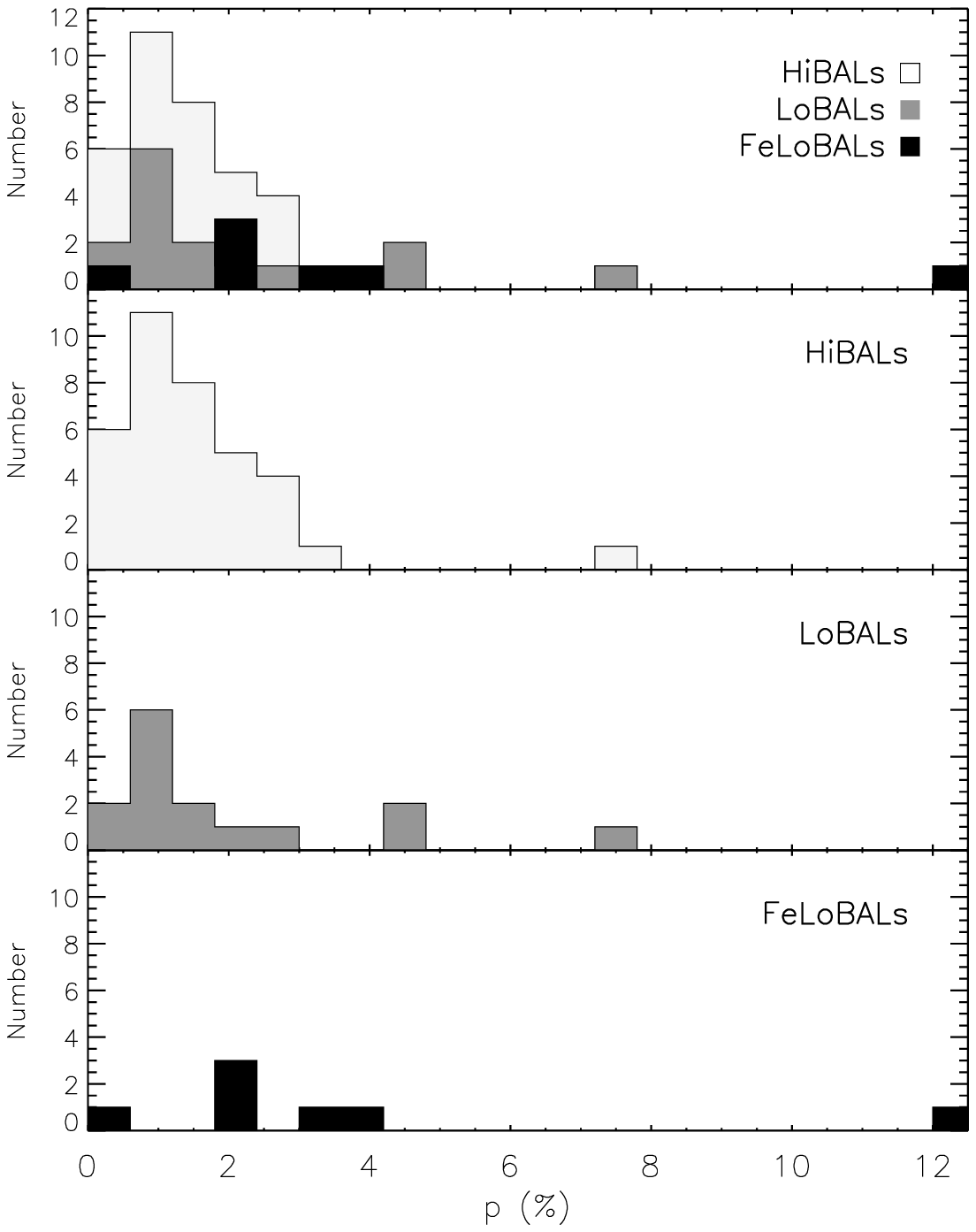}
 \caption{The distributions of continuum polarization, separated by BAL subclass.}
 \label{poldistfig}
\end{figure}

Of the 58 total sources, 18 of them have a high continuum polarization (defined here as $>$2\%).  This corresponds to 31.0\%$\pm$6.1\%, where the uncertainty is calculated from a binomial distribution.  This is consistent with previous findings, where around one-fourth to one-third of BAL quasars are highly polarized.  Broken down into subclasses, 8 of 36 (22.2\%$\pm$6.9\%) 
HiBAL quasars are highly polarized, compared to 10 of 22 (45.5\%$\pm$10.6\%) FeLo/LoBALs. This would seem to indicate that more reddened BAL quasars are more polarized, as generally FeLo/LoBALs are slightly more red (e.g. Sprayberry \& Foltz 1992, Brotherton et al. 2001, DiPompeo et al. 2012a).  Indeed, the most reddened sources in this sample are also Lo/FeLoBALs (e.g. see Figure~\ref{corrfig}b).  We show the continuum polarization distributions by BAL subtype in Figure~\ref{poldistfig}.  Note that despite the higher likelihood of an FeLo/LoBAL quasar being highly polarized, it seems that HiBALs and LoBALs can reach similar levels of polarization.

\begin{figure*}
 \includegraphics[width=6in]{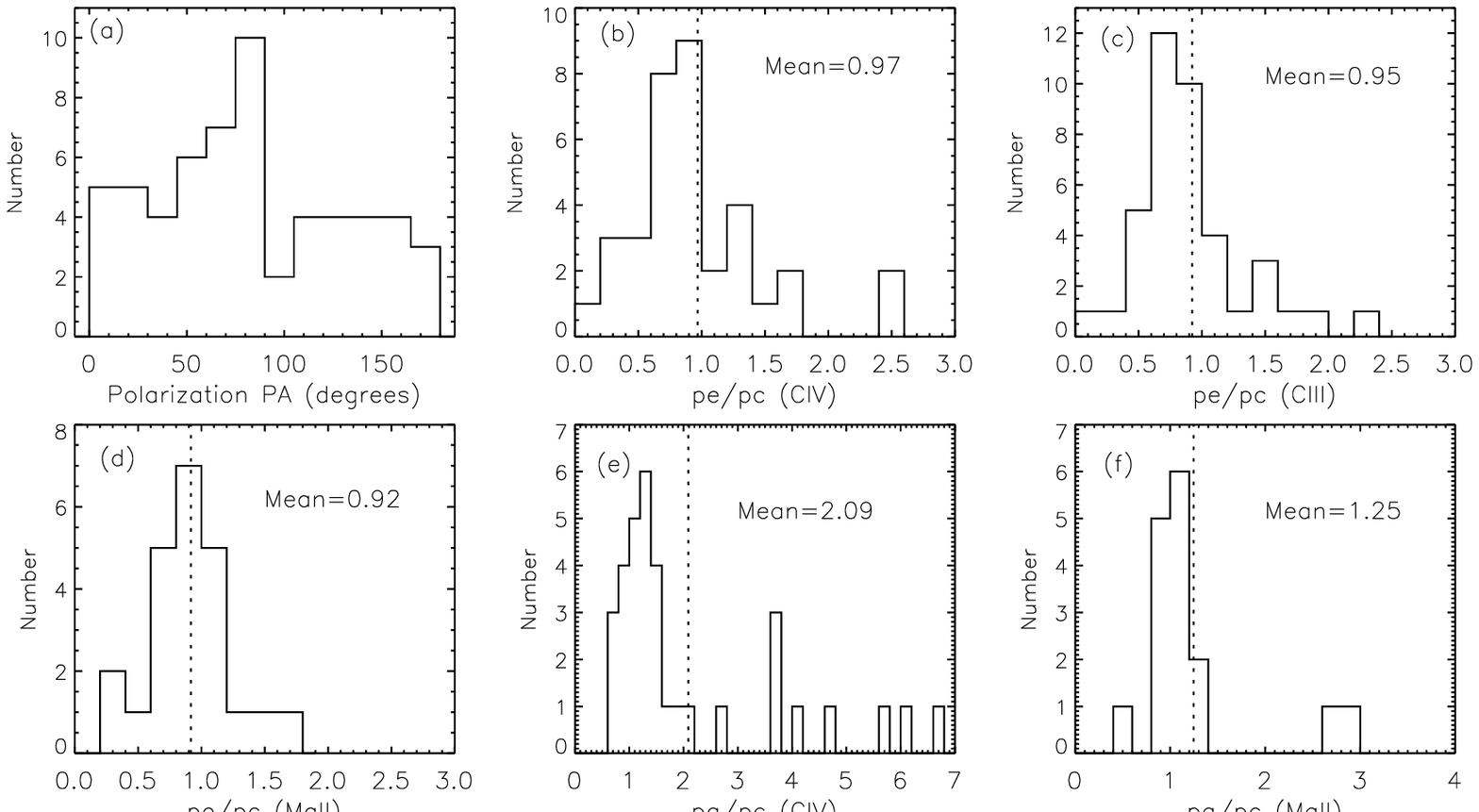}
 \caption{The distributions of other polarization properties (all line polarizations are normalized by the redward adjacent continuum): (a) The continuum polarization position angle. (b)-(d) The \CIV, \CIII, and \MgII\ emission line polarizations, respectively.  (e)-(f) The \CIV\ and \MgII\ absorption trough polarizations, respectively.  Dotted lines mark the distribution means.}
 \label{linepoldistfig}
\end{figure*}

\begin{figure*}
 \includegraphics[width=6.75in]{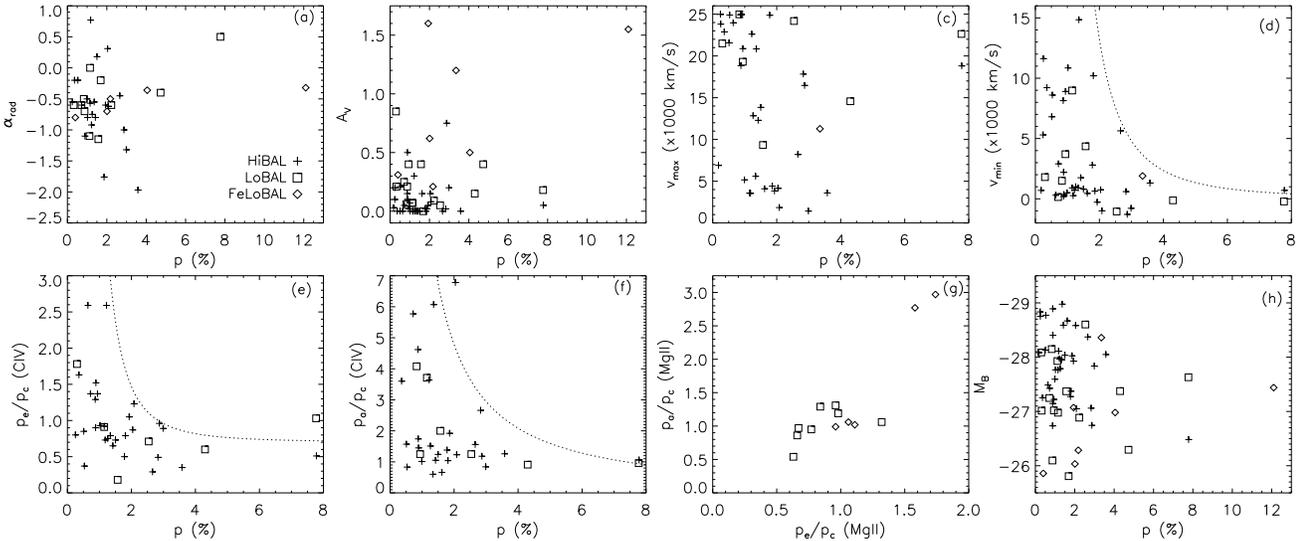}
 \caption{Visualizations of the significant correlations shown here, as well as some important non-correlations. From top left: (a) The non-correlation between $p$ and $\alpha_{rad}$.  (b) The non-correlation between $p$ and $A_V$.  (c) The correlation between $p$ and $v_{max}$.  (d) The correlation between $p$ and $v_{min}$, more of an upper envelope.  (e) The correlation between $p$ and the \CIV\ emission line polarization, also an upper envelope.  (f) The correlation between $p$ and the \CIV\ absorption trough polarization, another upper envelope.  (g) The correlation between the \MgII\ emission line and absorption trough polarization.  (h) Once corrected for reddening, there is no correlation between $M_B$ and $p$.  Note that in all panels, pluses denote HiBALs, squares denote LoBALs, and diamonds denote FeLoBALs.  The dotted lines in panels (d), (e) and (f) are empirically determined envelopes and are there to guide the readers eye.  All are simple power laws.}
 \label{corrfig}
\end{figure*}

Figure~\ref{linepoldistfig} shows distributions of the other polarization parameters.  Panel (a) shows the polarization position angle distribution, which is relatively flat as expected.  There may be a slight overabundance of sources with a position angle around 90$^{\circ}$, but given the numbers it is not overly convincing.  The next five panels, (b)-(f), show the polarization in various emission or absorption features; $p_e$/$p_c$ for \CIV, \CIII, and \MgII, respectively, and $p_a$/$p_c$ for \CIV\ and \MgII, respectively.  The means of each parameter are also shown in the figure.  It is clear that on average, broad emission lines are polarized at a level slightly less than the continuum, but there are many clear exceptions to this.  BAL troughs are generally more polarized than the continuum, but again there some objects where this is not the case.  Statistically, using a Wilcoxon signed-rank-sum test, we can only say with certainty that $p_a$/$p_c$ (\CIV) does not have a median of and is not symmetric about 1 ($p=6.4\times10^{-5}$).  Similar tests for all other line polarizations give $p\sim0.1$, though for $p_e$/$p_c$ (\CIII) it is lower at $p=0.03$

We have also performed an extensive search for correlations between various quasar properties and polarization, as well as between polarization properties, using the Spearman rank correlation coefficient ($r_s$).  These results are tabulated in Table~\ref{corrtbl}.  The first two columns list the two properties examined, the third column gives $r_s$, the fourth column gives the two-tailed probability that the test statistic is significantly different from zero ($P_{r_s}$), and the final column gives the number ($n$) of objects used in the test.  We use a cutoff of $P_{r_s} \leq 0.02$ to identify significant correlations; these values are typeset in bold in the table.  With 29 tests performed, we expect fewer than one correlation to appear by chance given our significance cutoff.  There are two particularly interesting non-correlations, which are shown in panels (a) and (b) of Figure~\ref{corrfig}.  Continuum polarization does not seem to correlate with $\alpha_{rad}$ nor the amount of reddening.  These will be discussed further in section 5.

We identify 5 significant correlations, which are shown graphically in Figure~\ref{corrfig}:
\begin{enumerate}
\renewcommand{\theenumi}{(\arabic{enumi})}

\item An anti-correlation between polarization and the maximum velocity of the \CIV\ BAL outflow (Figure~\ref{corrfig}c).  The sources with large polarization seem to be outliers however, and serve to weaken the correlation.  Better coverage of very high polarization sources may then make this correlation disappear.

\item An anti-correlation between polarization and $v_{min}$ of the \CIV\ BAL outflow (Figure~\ref{corrfig}d).  This does appear to be an upper envelope rather than a true correlation, and is analogous to the finding by Lamy \& Hutsemekers (2004) that $p$ and the detachment index ($DI$) form an envelope.

\item An anti-correlation between continuum polarization and the polarization in the \CIV\ broad emission line (Figure~\ref{corrfig}e).  Again, the relationship appears to form an envelope.

\item The same as above, but for the \CIV\ absorption trough polarization (Figure~\ref{corrfig}f).

\item A positive correlation between the polarization in the \MgII\ emission line and absorption trough (Figure~\ref{corrfig}g).  Note that this correlation is only based on 13 points and is strongly driven by the two sources with large values of $p_e$/$p_c$ and $p_a$/$p_c$.  Without these sources, the significance of the correlation drops below our cutoff, to $P_{r_s}=0.038$.  Because of this, we will not consider this correlation further until it is confirmed with more data.
\end{enumerate}

We realize that it may appear that some of these correlations are driven by the few objects with large polarization.  However, in all cases excluding these objects makes the already significant correlations stronger, due the the fact that they appear less like envelopes and more like true correlations.  On the other hand, none of the non-significant correlations become significant with the exclusion of these objects.

Briefly, we compare these results to some of the similar tests done by Lamy \& Hutsemekers (2004).  As stated above, we confirm their finding of an upper envelope in the $p$-$v_{min}$ relationship (they use $DI$, but this is very similar to $v_{min}$).  While we do not perform detailed spectral fitting here, we have done so for the VLA sample, which is presented in DiPompeo et al. (2012b).  We did not identify any correlation between $BI$ and \FeII\ strength (also seen by Weymann et al. 1991), nor between $BI$ and the slope of the ultraviolet continuum.  It is possible that these discrepancies point to a difference between radio-loud and radio-quiet BAL quasars.  We confirm that there is no correlation between $p$ and $BI$, and therefore the relationship found by Schmidt \& Hines (1999) is probably not real.  However, we caution this could be due to the large number of radio sources in our sample, and could signify a difference between radio-selected and radio-quiet BAL sources.  We also do not see the correlations between polarization in the \CIV\ trough and $BI$, or between $p_e$/$p_c$ in \CIV\ and \CIII.  Finally, we had seen a correlation between $M_B$ and polarization with the Keck and VLT samples analyzed individually.  However after more careful consideration of the effects of reddening here, this correlation disappears (Figure~\ref{corrfig}h).

\setcounter{table}{3}
\begin{table*}
  \caption{Correlation analysis of polarization properties.}
  \label{corrtbl}
  \begin{tabular}{ccccc}
  \hline
   Property 1 & Property 2 & $r_s$ & $P_{r_s}$ & $n$ \\
   \hline           
  Cont. $p$                          & $\alpha_{rad}$      &  0.034  & 0.836             & 38 \\
  Cont. $p$                          & $A_V$ ($A_g$)     &  0.101 (0.166)  & 0.448 (0.211)   & 58 \\
  Cont. $p$                          & $L_r$                      &  0.170  & 0.287              & 41 \\
  Cont. $p$                          & $R^*$                      &  0.282  & 0.074             & 41 \\
  Cont. $p$                          & $M_B$     	        &  0.200  & 0.132              & 58 \\
  Cont. $p$                          & $BI$ ($AI$)             & 0.019 (0.022)   & 0.916 (0.890)      & 32 (41) \\
  Cont. $p$                          & $v_{max}$              & -0.479  & \textbf{0.001} & 41 \\
  Cont. $p$                          & $v_{min}$               & -0.421  & \textbf{0.004} & 46 \\
  Cont. $p$                          & $p_e/p_c$(\CIV)    &  -0.465 & \textbf{0.005} & 35 \\
  Cont. $p$                          & $p_e/p_c$(\CIII)     &  0.143 & 0.379              & 40 \\
  Cont. $p$                          & $p_e/p_c$(\MgII)   & -0.092 & 0.676              & 23 \\
  Cont. $p$                          & $p_a/p_c$(\CIV)    & -0.407 & \textbf{0.017} & 34 \\
  Cont. $p$                          & $p_a/p_c$(\MgII)   & -0.086 & 0.749              & 16 \\      
  $BI$ ($AI$)                       & $p_e/p_c$(\CIV)   & 0.265 (0.238)   & 0.200 (0.195)     & 25 (31) \\ 
  $v_{max}$                        & $p_e/p_c$(\CIV)   & 0.241    & 0.192             & 31 \\ 
  $v_{min}$                         & $p_e/p_c$(\CIV)   & -0.058   & 0.741             & 35 \\ 
  $BI$ ($AI$)                       & $p_a/p_c$(\CIV)   & 0.069 (-0.045)  & 0.734 (0.807)     & 26 (31) \\ 
  $v_{max}$                        & $p_a/p_c$(\CIV)   &  0.317    & 0.082             & 31 \\ 
  $v_{min}$                         & $p_a/p_c$(\CIV)   &  0.391    & 0.022             & 34 \\ 
  $p_e/p_c$(\CIV)              & $p_e/p_c$(\CIII)    &  0.299 & 0.108               & 30 \\                     
  $p_e/p_c$(\CIV)              & $p_e/p_c$(\MgII)   & 0.445 & 0.169               & 11 \\ 
  $p_e/p_c$(\CIV)              & $p_a/p_c$(\CIV)    &  0.305 & 0.122               & 27 \\
  $p_e/p_c$(\CIV)              & $p_a/p_c$(\MgII)   & -0.200 & 0.747               & 5  \\
  $p_e/p_c$(\CIII)               & $p_e/p_c$(\MgII)   &  0.251 & 0.349               & 16 \\
  $p_e/p_c$(\CIII)               & $p_a/p_c$(\CIV)    &  0.289 & 0.144               & 27 \\
  $p_e/p_c$(\CIII)               & $p_a/p_c$(\MgII)   &  0.252 & 0.513                &  9 \\
  $p_e/p_c$(\MgII)             & $p_a/p_c$(\CIV)    & -0.130 & 0.703               &  11 \\
  $p_e/p_c$(\MgII)             & $p_a/p_c$(\MgII)   &  0.776  & \textbf{0.002} & 13 \\
  $p_a/p_c$(\CIV)              & $p_a/p_c$(\MgII)   & -0.300 & 0.624               &  5 \\
   \hline
   \end{tabular}
   \\
   {
   \raggedright
   The first two columns give the parameters being compared.  $r_s$ is the Spearman rank correlation coefficient, and $P_{r_s}$ is the two-sided significance of its deviation from 0.  Values of $P_{r_s}$ smaller than 0.02 are highlighted in bold.  $n$ is the number of sources included in each test.\\
   }
   \end{table*}

We pointed out earlier that the VLT sample is heterogeneous, in that it consists only of particularly bright and often extreme examples of BALs, and therefore we have also performed the analysis with this subsample excluded to ensure that our results are not biased in some way (such as being driven by the brightest sources).  The remaining sample is homogeneous, in that it is simply composed of radio-loud BAL quasars with a range of other properties.  Without the VLT sample, in general, the above results hold.  The percentages of highly polarized BAL quasars, including when only considering Hi or LoBALs, do not change significantly.  The trends seen in emission and absorption line polarizations remain the same as well.  The significance of the four correlations seen with continuum polarization ($v_{max}$, $v_{min}$, $p_e$/$p_c$ (\CIV), and $p_a$/$p_c$ (\CIV)) do drop below our significance cutoff, but this is mostly due to their appearance as upper envelopes rather than true correlations.  Visually, the relationships look quite similar.  We do however see the emergence of a significant correlation between $BI$ and the polarization in the \CIV\ emission line, with $P_{r_s}=0.01$ (and $n=14$).  However, there are quite a few large outliers, and as there is no clear explanation of this relationship we believe it may not be real.  No other correlations become significant with the exclusion of the VLT sample.

\section{DISCUSSION}
One of the primary motivations of this study was to look for a correlation between $\alpha_{rad}$ and continuum polarization.  It is now very likely that BAL quasars are seen along a wide range of viewing angles, but this does not mean that the number of high-polarization sources is not related to orientation.  If the most polarized sources are indeed seen edge-on, then as $\alpha_{rad}$ decreases (becomes more negative) the polarization should increase.  As shown in panel (a) of Figure~\ref{corrfig}, there is no correlation, suggesting that there is no strong relationship between viewing angle and polarization.  This seems to rule out models where the polarization is only due to an edge-on view of equatorial BAL winds with a polar scattering region.

One possible caveat is that the spectral index distribution for this sample happens to be clustered around $-$0.5.  This region of parameter space is somewhat ambiguous with regards to viewing angle due to the amount of intrinsic scatter in the $\alpha_{rad}$-viewing angle relationship (DiPompeo et al. 2012a, for example).  However, the range in $\alpha_{rad}$ for this sample does cover the full range of observed spectral indices, from around $-$2 to 1.  Additionally, the four most highly polarized sources ($p>4$\%) would be considered flat spectrum sources.  We see this as a clear indication that a BAL source need not be seen edge-on to be highly polarized.

Another simple explanation for the polarization in BAL quasars could be related to their higher amounts of reddening.  If the direct, unpolarized continuum light from the accretion disk is attenuated by the dust in the wind, then the ratio of scattered to direct light in BAL sources is higher than in non-BAL sources, leading to higher polarizations on average.  The discovery of objects like FIRST J1556+3517 seemed to support this view, as it is one of the most highly polarized and highly reddened BAL quasars known.  The percentages of highly polarized sources by subclass found here also seem to support this view.  It is known that FeLo/LoBAL quasars are more reddened than HiBAL quasars (notice in panel (b) of Figure~\ref{corrfig} that the most reddened sources are almost all FeLo/LoBALs; see also DiPompeo et al. 2012b), and they are nearly twice as likely to have polarizations above 2\%.  

However, Lamy \& Hutsemekers (2004) saw no correlation between $p$ and $\alpha_{uv}$, which is a fairly good indicator of the amount of reddening.  We confirm that here with our measurement of $A_V$ (and $A_g$), as shown in panel (b) of Figure~\ref{corrfig}.  Clearly there are sources with a significant amount of reddening that are not highly polarized.  Likewise, there are sources without significant reddening that appear highly polarized, particularly the two sources with very high polarization (nearly 8\%) and values of $A_V$ only around 0.1-0.2.  Some have speculated that BAL quasars are intrinsically bluer than non-BAL quasars (e.g. Reichard et al. 2003), and so it is possible that the amount of reddening measured here is actually an underestimation.  However, unless the intrinsic continuum shape in BAL quasars depends strongly on some other fundamental parameter (i.e. it can vary significantly between sources), this should simply have a systematic effect on our measurements and no effect on the correlation analysis.  Thus, it seems that dust reddening, like orientation, is insufficient to explain the polarization in BAL quasars.  This is also supported by the lack of a correlation between polarization and $R^*$.

As previously mentioned, we do confirm that on average broad emission lines (BELs) are less polarized than the surrounding continuum (or at a level similar to the continuum).  This is generally interpreted to mean that the BEL region lies either coincident with or outside the scattering region.  This may be the case in many objects, however for all emission lines there are a significant number of sources where the polarization does increase.  In fact, as shown in Figure~\ref{linepoldistfig} (panels b-d), the distributions of $p_e$/$p_c$ appear approximately Gaussian around 1.  This seems to indicate that there is significant variation in the scatterer location relative to the BEL region.

The BAL troughs, on the other hand, seem to be almost always more polarized than the surrounding continuum, and the distributions in Figure~\ref{linepoldistfig} do not appear normal as they do for the BELs.  Obviously there is significant blocking of the direct continuum light in the BALs, and so this naturally leads to an increase in polarization and supports the idea that the polarization is indeed to due to scattering well outside the BAL region.  The polarization increase in the \CIV\ BALs is generally stronger than in the \MgII\ BALs, likely because the \CIV\ troughs are often much deeper than the \MgII\ troughs, and in some cases almost completely black.  An extreme example of this is seen in the spectropolarimetry plots for 1159+0112 in Figure~\ref{polfig} (d), where the polarization jumps to near 14\% in the very deep \CIV\ BAL (as well as the \NV\ BAL).

Spectropolarimetry can be particularly useful for sources with resolved radio jets, because it allows a comparison between the radio jet position angle and the polarization position angle.  With a more edge-on view and scatterers in a polar region (maybe even in the jet itself), we expect these position angles to be roughly perpendicular.  Alternatively, a more pole-on view with an equatorial scattering region will have these position angles parallel.  Indeed, the former is generally the case, for both quasars and type 2 AGN (di Serego Alighiere et al. 1993, Vernet et al. 2001, Brotherton et al. 2006, DiPompeo et al. 2010, 2011a), but the numbers are quite small.  Unfortunately, none of the newly observed sources presented here are resolved by FIRST or our own observations and so we cannot apply this test.  Since our other results seem to rule out a picture in which BAL quasar polarization is due to orientation alone, it may simply be that the sources sufficiently resolved enough to see radio jets are generally seen edge-on and the scattered light (which may occur in multiple regions) is dominated by that from the polar regions.

Lamy \& Hutsemekers (2004) identified the correlation between polarization and $DI$ as potentially one of the more important ones, and argued that it had a natural explanation in orientation (see their Figure 8).  While we confirm this correlation here via our measurement of $v_{min}$, and do not deny that their interpretation seemed reasonable, it is difficult to accept given some of the new findings here (and other recent BAL studies).  In particular, the fact that more evidence is mounting that polar BALs exist and there is no correlation between polarization and $\alpha_{rad}$ seems to rule out this simple picture.  Also, we do not see a correlation between $\alpha_{rad}$ and $v_{min}$, neither with this sample nor with the VLA sample.  

One possibility is that some BAL outflows are launched in bursts, while others are more continuous in nature, regardless of which direction they are launched in, and polarization is indeed due to scattering off the winds themselves.  In this case, in a continuous flow through which we see all velocities (small $DI$ or $v_{min}$), there is sufficient material for light to scatter from and polarization may or may not be seen, depending on the projected scattering geometry.  Thus, sources with small $v_{min}$ can have a range of polarizations.  In sources where BAL clouds are launched in a burst and we only see the absorption after the cloud is accelerated to high velocity (large $v_{min}$), there isn't much material behind it to scatter continuum light, and so large $v_{min}$ sources are always low polarization.

The anti-correlation/upper envelope in the \CIV\ absorption and continuum polarizations has a relatively natural explanation.  Objects in the upper left of this plot probably have scattering occurring outside the absorbing outflow, such that the scattered light is not absorbed and we see a high polarization in the absorption troughs where direct light is severely attenuated.  Objects in the lower half of the plot likely have both scattered and direct light being attenuated, indicating that the scattering occurs interior to the absorbing region causing the polarization in the line and continuum to be similar.  The relative locations of the scatterers and absorbing wind cause the scatter around one here.  The region devoid of any objects is probably due to the amount of polarization possible in objects seen at these viewing angles.  For example, an object with a continuum polarization of about 4\% would need to have a polarization of above 12\% in the absorption trough to even begin to fall above the upper envelope seen.  For an object with 8\% continuum polarization, it would need to have polarization in the line above around 15\%.  Since we know these objects are not seen beyond about 40$^{\circ}$ from the symmetry axis in the most extreme cases (DiPompeo et al. 2012a), modeling has shown that these polarizations become difficult to achieve even if all of the direct light is absorbed while the scattered light is not absorbed at all (Brotherton et al. 1998).
  
The anti-correlation between continuum polarization and the polarization in the \CIV\ emission line is more difficult to interpret.  It is likely that the zone of avoidance here has a similar explanation as above.  However, the absorption of direct and/or scattered light is required to explain the other regions, so it cannot be used to explain this relationship with emission lines.

All of these results seem to paint a picture in which the relatively high polarization of BAL quasars could be due to a multitude of factors, and one explanation such as an edge-on view is simply insufficient.  Clearly BAL quasars along any line of sight can be highly polarized, and BAL quasars with various amounts of reddening can have similar polarizations.  The two-component wind model proposed by Lamy \& Hutsemekers may have some merit, with the exception that the polar component is not necessarily less dense than the equatorial one, since it seems that BAL outflows along any line of sight have similar properties (DiPompeo et al. 2012b).  It is likely that there are multiple scattering paths, evidenced by the results above as well as the rotation of the polarization position angle across absorption features seen in many objects, and it could be that the winds themselves are the scatterers.

\section{SUMMARY}
We have presented new spectropolarimetric observations of 8 BAL quasars from the sample of DiPompeo et al. (2011b), and combined these measurements with the observations presented by DiPompeo et al. (2010, 2011a) in order to analyze the polarization of BAL quasars as a group.  Many of our objects are radio selected, and have radio spectral indices available as an orientation indicator.  We find that about 30\% of BAL quasars are polarized at a level greater than 2\%.  When split into subclasses, FeLo/LoBALs are about twice as likely to be highly polarized.  However, we also find that polarization does not correlate with either radio spectral index or the amount of intrinsic reddening, despite the fact that FeLo/LoBALs are redder on average.  This suggests that the polarization in BAL quasars cannot be explained by a preferred viewing angle or the blocking of direct unpolarized continuum light alone.  While these could both be important factors, or some other as yet unknown mechanism, it seems there is significant variation between sources and one mechanism may dominate for some objects while another dominates for others.

We also find that while on average the polarization in broad emission lines is lower (or similar to) the continuum, there appears to be a smooth distribution with a non-negligible number of sources not following this trend.  This is another indication of significant variation among sources in the scatterer location and geometry.  BAL troughs on the other hand, are almost always more polarized than the surrounding continuum, with few small exceptions.  

We identify a few correlations between various quasar properties and polarization, as well as between some polarization properties.  There is an inverse correlation (really more of an upper envelope) between continuum polarization and the minimum BAL outflow velocity, confirming results from other groups.  A similar relationship may exist with the maximum outflow velocity of the  BAL trough, though it is of lower significance and there are a few questionable outliers.  Both the polarization in the \CIV\ emission line and absorption trough decrease with continuum polarization.

While a simple orientation-dependent picture for the polarization in BAL quasars similar to what is found for Seyfert type galaxies is nice in principle, it appears that it does not work in practice.  Moreover, it appears that there is significant variation between sources in scatterer location and geometry, which may make it difficult to use spectropolarimetry as a tool to analyze BAL quasars as a group, though it is potentially useful for detailed study of individual objects.

\section*{acknowledgments}
M. DiPompeo thanks the European Southern Observatory for awarding two DGDF fellowships used to fund visits to ESO headquarters, which provided fruitful face-to-face collaboration with C. De Breuck.  M. DiPompeo also thanks the Wyoming Nasa Space Grant Consortium for providing funding to complete a portion of this work.

\label{lastpage}

\end{document}